\documentclass{article}

\usepackage{amssymb}
\usepackage{verbatim}
\usepackage{lscape}

\author{Aron C. Wall\footnote{aronwall@umd.edu}
\\ \textit{Maryland Center for Fundamental Physics} \\ \textit{Department of Physics} \\ \textit{University of Maryland} \\ \textit{College Park, MD 20740-4111, USA} }
\title{Ten Proofs of the Generalized Second Law}
\date{\today}

\begin{document}

\maketitle

\begin{abstract}
Ten attempts to prove the Generalized Second Law of Thermodyanmics (GSL) are described and critiqued.  Each proof provides valuable insights which should be useful for constructing future, more complete proofs.  Rather than merely summarizing previous research, this review offers new perspectives, and strategies for overcoming limitations of the existing proofs.  A long introductory section addresses some choices that must be made in any formulation the GSL: Should one use the Gibbs or the Boltzmann entropy?  Should one use the global or the apparent horizon?  Is it necessary to assume any entropy bounds?  If the area has quantum fluctuations, should the GSL apply to the average area?  The definition and implications of the classical, hydrodynamic, semiclassical and full quantum gravity regimes are also discussed.  A lack of agreement regarding how to define the ``quasi-stationary'' regime is addressed by distinguishing it from the ``quasi-steady'' regime.
\end{abstract}

\newpage
\tableofcontents
\newpage

\section{Introduction}\label{intro}

In this review I summarize and critique several attempts to prove the Generalized Second Law (GSL).  Here a ``proof'' means a detailed argument trying to establish the GSL for a broad range of states in some particular regime.  Thus I do not include results showing that the second law holds in some particular state.  Disregarding chronology, I have classified the proofs based on the core concepts used.

Most of the proofs are unsound.  Some have inconsistent or erroneous assumptions, and others have hidden gaps in the reasoning.  Nevertheless each of these proofs is valuable.  Even an invalid proof can clarify the issues and choices that must be resolved in order to fully understand the GSL.  Faulty proofs might also be correctable through small adjustments.  It is better to view them as research programs than as mere fallacies.

\subsection{What does the Generalized Second Law say?}\label{def}

The Ordinary Second Law (OSL) states that the total thermodynamic entropy of the universe is always nondecreasing with time.  In a background-free theory such as General Relativity (GR), a ``time'' is a complete spatial slice, and a ``later time'' is a complete slice which is entirely in the future of the earlier time slice.

The GSL states that the ``generalized entropy'' of the universe is nondecreasing with time.  This generalized entropy is given by the expression
\begin{equation}\label{Gen}
\frac{kA}{4G \hbar} + S_{out},
\end{equation}
where $k$ is Boltzmann's constant, $c = 1$ \cite{hawking75},\footnote{After section \ref{intro}, I will normally use $k$ = $\hbar$ = $G$ = 1.} and $A$ is the sum of the area of all black hole horizons in the universe, while $S_{out}$ is the ordinary thermodynamic entropy of the system outside of all event horizons.  The first term is called the Bekenstein-Hawking entropy ($S_{BH}$).  Since the horizon area and the outside entropy are time-dependent quantities, each term is defined (like the ordinary entropy) using a complete spatial slice.

The above description is still very imprecise; there are several ways to interpret it.  The first step towards a proof must be to give a definition of the generalized entropy above.

\subsubsection{Boltzmann or Gibbs?}\label{borg}

Even in ordinary thermodynamics there are multiple ways to define the ``entropy'' \cite{jaynes65}.  The ``Boltzmann entropy'' requires a choice of coarse-grained observables capable of being measured macroscopically.  A ``macro\-state'' is then a class of $N$ pure states all having the same values of all coarse-grained observables.  Then each pure state in the class has entropy given by 
$S = k\ln N$.  One then tries to prove the OSL by showing that typical states in a macrostate are unlikely to evolve to  another macrostate with much smaller $N$ value, but might evolve to a microstate with much larger $N$ value.  Since the ratios of $N$ values are typically huge in standard thermodynamic applications, the Boltzmann entropy of a typically prepared low-entropy state nearly always increases in entropy over time, except for small fluctuations.  (However, if the state were truly typical the argument could be reversed to show that the entropy also increases in the past direction.  Thus a real proof must also show that states which are atypical in the sense that they have low entropy pasts are still sufficiently ``typical'' for purposes of future evolution.)  For a fully quantum mechanical discussion of the Boltzmann entropy see Wald \cite{wald79}.

Another choice is the ``Gibbs entropy'', which assigns an entropy to mixed states.  A probability mixture over $N$ states has entropy
\begin{equation}
S = k\sum_i -p_i\ln p_i.
\end{equation}
This definition does not yet require any notion of coarse-graining.  It agrees with the Boltzmann entropy in the case of a uniform mixture over all the pure states in a single macrostate.  The generalization to a quantum state with density matrix $\rho$ is
\begin{equation}
S = -k\,\mathrm{tr}(\rho \ln \rho).
\end{equation} 
This entropy is conserved under unitary time evolution.  This means that the OSL is trivially true for an ordinary closed quantum mechanical system, away from any black holes.  A real proof of the OSL using the Gibbs entropy must also explain why entropy seems to increase.\footnote{A Bayesian might propose that any observer who does not know the exact Hamiltonian of a system should predict the future using a probability distribution over the possible unitary evolution rules.  This coarse-grained evolution rule will turn pure states into mixed states.  But since every unitary evolution rule preserves the maximum entropy state, a mixture of different unitary evolution rules also preserves the maximum entropy state.  Theorem 1 from section \ref{special} then implies the OSL.}

The Gibbs entropy does not fluctuate about its maximum value like the Boltzmann entropy does.  Hence the Gibbs definition is more convenient for proofs because it allows one to state without reservation that the entropy of the state always increases with time.  Presumably this is why all proofs below except one use the Gibbs entropy.  The exception is Fiola et al. \cite{fiola94} (section \ref{2D}), which combines the Gibbs and Boltzmann concepts (cf. section \ref{Boltz}).

The choice between Gibbs and Boltzmann also has implications for the interpretation of the area component of the generalized entropy.  Consider a black hole in a mixed state which has different possible values of the $A$, but has fixed $S_{out}$.  Should one say that the mixed state has an uncertain entropy?  Or should one simply calculate the entropy using the expectation value of the area?  The former choice seems to be analogous to the Boltzmann approach, since entropy values only to pure states, leading to statistical fluctuations in the entropy even in equilibrium.  The latter choice is more like the Gibbs approach since the entropy is a function of a mixed state $\rho$.  By taking the Gibbs approach to both terms in the generalized entropy, one ends up with a simple trace formula for the generalized entropy:
\begin{equation}
S = k\,\mathrm{tr}(\rho\,(A - \ln \rho)) = k \left(\frac{\langle A \rangle}{4G \hbar} - 
\mathrm{tr}(\rho\ln \rho) \right).
\end{equation}
The use of the expectation value of the entropy in situations where there are fluctuations in the area is further supported by arguments in Ref. \cite{SS99}.

There are some respects in which proving the GSL is easier than proving the OSL.  For example, the black hole horizon favors one direction of time by definition, removing the problem of getting a time asymmetric result from time symmetric assumptions.  And unlike the ordinary entropy, the generalized entropy does not require an arbitrary method of coarse graining to get an entropy increase, since the horizon determines what is observable outside in an objective way \cite{sorkin05}.  Under this understanding, the generalized entropy at one time does not depend on any details about the time slice except where the slice intersects with black hole horizons.

\subsubsection{The Choice of Horizon}\label{choice}

The GSL seems to apply not only to black hole horizons, but also to de Sitter and Rindler horizons.  Arguably the only requirement is that the horizon be the boundary of the past of some infinite worldline \cite{JP03}.  However, the GSL cannot apply to every null surface.  For example, consider a trapped spherically symmetric surface well inside the horizon of a Schwarzschild black hole.  Take the quantum field theory in curved spacetime limit: $G \to 0$ while holding the black hole radius $R$ constant.  Since the area of such a trapped surface decreases even classically, the total decrease in the entropy is of order $G^{-1}$ due to the $G$ in the denominator in Eq. (\ref{Gen}).  This decrease cannot be atoned for by an increase in the $S_{out}$ term, because this term is finite in the quantum field theory limit and thus has no scale dependence on $G$.

Conventional wisdom suggests that the GSL should hold on the global event horizon, i.e. the boundary of the past of $\mathcal{I}^{+}$.  This is defined by a ``teleological'' boundary condition, meaning that the location of the boundary at one time can depend on what will happen later in time \cite{PTR86}.  The event horizon is defined using the causal structure, a more primitive concept than the metric, and therefore more likely to be meaningful in a full quantum gravity theory.  The event horizon is always a null surface, appropriate to the thermodynamic role it plays as a concealer of information, while the apparent horizon may be spacelike or timelike depending on the dynamics of the situation.  Furthermore the location of the apparent horizon, since it is local, is more sensitive to metric fluctuations, so the event horizon is more likely to be well defined in full quantum gravity \cite{SS99}.

Nevertheless, analogues of the classical laws of black hole mechanics have been proposed for the apparent horizon \cite{hayward93}, and some suggest that the GSL should apply to the apparent horizon, defined as a marginally trapped surface around the black hole \cite{ZWGA07}.  Unlike the event horizon, the apparent horizon is sometimes spacelike or timelike and thus it sometimes permits information to escape.  The only proof reviewed here which uses the apparent horizon is that of Fiola et al. \cite{fiola94}.  Their argument for the apparent horizon is discussed in section \ref{horizon}.

\subsection{Types of Regimes}\label{regimes}

The interpretation of the generalized entropy also depends on which regime a proof is set in, i.e. what restrictions the proof needs to impose on the perturbations of the black hole.

The first question is how large and how rapidly changing these perturbations are allowed to be (sections \ref{quasi}-\ref{adiabatic}).

The second question is how many features of quantum mechanics are taken into account.  The answer to this will determine whether the proof is set in the classical, hydrodynamic, semiclassical, or full quantum gravity regimes (sections \ref{class}-\ref{QG}).  Each of these four regimes involves a different interpretation of the exterior entropy term $S_{out}$.

\subsubsection{The Quasi-stationary and Quasi-steady Regimes}\label{quasi}

This section describes two distinct regimes.  Confusingly, each has been called the ``quasi-stationary'' regime by different authors.  I will suggest that one regime should retain the name, while for the other I propose the name ``quasi-steady''.

For example, Sorkin uses the term ``quasi-stationary'' to mean that
\begin{quote}\small
[...] we assume that the spacetime geometry can be well approximated at any stage by a strictly stationary metric. [...] Notice that the requirement of approximate stationarity applies only to the metric; the matter fields (among which we may include gravitons) can be doing anything they like.  [I have used the ellipses here to disentangle this definition from Sorkin's commingled definition of ``quasi-classical''.] (\cite{sorkin98} p. 12) \end{quote}
Here the term ``quasi-stationary'' refers to any small, but otherwise arbitrary, perturbation to a stationary background metric.  This requires that the black hole radius satisfy $R \gg L_P$, or else the Hawking radiation coming from the black hole will itself be a large perturbation.  I will be using this definition of ``quasi-stationary'' in this review.

Frolov and Page appear to be using a different definition when they state that:
\begin{quote}\small
One would conjecture that the generalized second law applies also for rapid changes to a black hole, but then $S_{BH}$, one-quarter of the horizon area, would depend upon the future evolution.  One would presumably also need to include matter near the hole in [$S_{out}$], but it is problematic how to do that in a precise way without getting divergences from infinitely short wavelength modes if there is to be a sharp cutoff to exclude matter inside the hole.  In a quasistationary process, one can with negligible error allow enough time for the modes to propagate far from the black hole, where the states $\rho_1$ and $\rho_2$ and their respective entropies can be evaluated unambiguously. (\cite{FP93} p. 3903)
\end{quote}
Here the same word is being used to mean that there are no rapid changes, so that one does not need to know the future state of matter to calculate $S_{BH}$.  This means that the state the matter fields are in is an approximately steady state with respect to the Killing field that generates the horizon, over periods of time on the order of the black hole radius $R$.  I will refer to this as the ``quasi-steady'' regime, because it requires the system to be in an approximately steady state.  The quasi-steady regime implies the quasi-stationary regime, because it makes no sense to talk about unchanging matter fields living on a changing metric.  But the converse does not follow, because it is possible for the power absorbed by a black hole to be small in magnitude but still rapidly changing with time.  As it happens though, all proofs reviewed here either permit large fluctuations (i.e. are \emph{not} quasi-stationary proofs) or else require the fluctuations to be slow as well as small (i.e. are quasi-steady proofs).

Note that in the quasi-steady regime, large changes in $S_{BH} \sim R^2/L_{P}^2$ are still permitted if they are caused by a nearly constant influx of energy into the black hole; the requirement that the perturbation to the metric be small only requires that
\begin{equation}\label{primus}
R \frac{dS_{BH}}{dt} \ll S_{BH} \sim \frac{R^2}{L_P^2}.
\end{equation}
On the other hand, the second derivative of $S_{BH}$ is related to the \emph{change} in the energy falling into the black hole, and is therefore required to be much smaller:
\begin{equation}\label{secundus}
R \frac{d^2 S_{BH}}{dt^2} \ll \frac{dS_{BH}}{dt}.
\end{equation}

\paragraph{The First Law}

The quasi-steady approximation is useful because it implies the First Law \cite{BCH73}\cite{GW01} of black hole mechanics, viewed as a relation which holds between arbitrary slices of the black hole event horizon \cite{SCD81}\cite{JP03}.  The background spacetime (about which these quasi-steady perturbations are made) is the Kerr-Newman electrovac solution to the Einstein field equations.

One must be careful in defining the notion of ``time translation'' because it depends on the choice of electromagnetic gauge.  To describe events distant from the black hole, it is most natural to use a gauge choice in which the connection $A_a$ vanishes at spatial infinity.  Since the Kerr-Newman spacetime is asymptotically Minkowskian, one can then identify the time-translation Killing vector $\xi_t$, rotational symmetry $\xi_{\phi}$, and the electromagnetic U(1) phase shift based on their action on the asymptotic region.  These generate conserved quantities: the Killing energy $E$, angular momentum $J$, and charge $Q$ respectively.  Using the quasi-steady approximation, it now follows that between any two slices of the perturbed black hole's event horizon,
\begin{equation}\label{FIRST}
dE = T\,dS_{BH} + \Omega\,dJ + \Phi\,dQ,
\end{equation}
where $dE$, $dJ$, and $dQ$ are the fluxes of Killing energy, angular momentum, and charge into the black hole between the two slices, $T$ is the Hawking temperature, $\Omega$ is the angular velocity and $\Phi$ is the electrostatic potential on the horizon. \cite{SCD81}\cite{GW01}.  (Since $E$, $J$, and $Q$ are conserved, the flux of these quantities into the black hole is equal to the change in the mass, angular momentum, and charge of the black hole itself.)

On the other hand, to describe events near the black hole's event horizon, it is more natural to use a different notion of time translation coming from the horizon generating Killing vector $\xi_H = \xi_t + \Omega \xi_{\phi}$.  It is also more natural to use a gauge choice in which the potential vanishes on the horizon (i.e. $A_a \xi^a_H|_{horizon} = 0$), rather than at asymptotic infinity.  The flow of $\xi_H$ is then a combination of asymptotic time-translation, rotation, and phase shifting.  The Killing `energy' generated by $\xi_{H}$ is
\begin{equation}
E^{\prime} = E - \Omega J - \Phi Q,
\end{equation}
which is proportional to the energy defined relative to a ``fiducial observer'' who co-rotates with the black hole near the horizon.  This permits the expression of the First Law in a more compact form:
\begin{equation}\label{first}
dE^{\prime} = T dS,
\end{equation}
which is the form that will be used in several of the proofs below.

In order to deduce Eq. (\ref{FIRST}), the quasi-steady regime must require that the state be slowly changing, not with respect to the $\xi_t$ Killing flow, but with respect to the $\xi_H$ \cite{JP03}.  Only in the ``quasi-static'' case where the background metric is a non-rotating black hole, are they the same.  For example, a rapidly rotating black hole illuminated continuously by light from the ``fixed stars'' is not quasi-steady, because the incoming starlight is stationary with respect to the wrong Killing field.  This restriction may seem pedantic, but it is necessary to derive the First Law (\ref{FIRST}) as applied to arbitrary slices of the horizon.  Since GSL as I have defined it in section \ref{def} also applies to arbitrary slices of the horizon, any proof of the GSL which uses the First Law as a step implicitly assumes the quasi-steady regime.\footnote{If only the quasi-stationary approximation holds, the First Law still applies when comparing the black hole before and after the perturbation is made.  But then it cannot be used to rule out temporary decreases of the entropy during the perturbative process, so one only gets a weaker form of the GSL.}

\subsubsection{The Adiabatic Limit}\label{adiabatic}

I will use the term ``adiabatic'' to refer to a process which is described by the time 
evolution of a first order deviation from the Hartle-Hawking equilibrium state $\rho_{HH}$.\footnote{Jacobson and Parentani \cite{JP03} use the term ``adiabatic'' to refer to what I am calling quasi-steady processes.  This is similar to the definition of ``adiabatic'' in mechanics, but I would like to reserve that term here for the thermodynamic meaning, to describe a process which is always near thermal equilibrium.}  This limit is arguably used by the proof in Wald \cite{wald94} (section \ref{atm}).

More precisely, given any state $\rho$, one can define a one-parameter family of states:
\begin{equation}
\sigma(\epsilon) = (1 - \epsilon) \rho_{HH} + \epsilon \rho.
\end{equation}
This is a positive density matrix, at least for $0 \le \epsilon \le 1$.  However, some quantities of thermodynamic importance---such as the entropy---are undefined except for positive states.  For these quantities one should not expect expect a Taylor series in $\epsilon$ to converge unless $\sigma(\epsilon)$ is also positive for small negative values of $\epsilon$.  Also, in a system with infinitely many degrees of freedom, there may exist states $\rho$ whose generalized entropy is infinitely less than that of the Hartle-Hawking state.  Assuming that $\rho$ is selected to avoid these pathologies, and that $\epsilon$ is a small parameter, the state $\sigma$ is adiabatic.

Assuming that the GSL is true, in the adiabatic limit all processes are reversible (in the sense that the generalized entropy is constant with time).  This is because $dS/dt$, viewed as a function of the state, takes its minimal value of zero in the Hartle-Hawking state, and must therefore be constant to first order as one departs from the Hartle-Hawking state.  Some examples of this are given in Ref. \cite{bek98}.

An adiabatic perturbation is even smaller than a quasi-stationary perturbation, because it is not only small in its gravitational effect on the background metric, but also small in its effect on the thermal atmosphere of the black hole.  Surprisingly, an adiabatic perturbation need not necessarily be quasi-steady.  If $\rho$ is a rapidly changing state, then $\sigma$
is an adiabatic state which is still rapidly evolving with time.  Thus the quasi-steady adiabatic regime is more restrictive than either regime taken separately.

\subsubsection{Classical Black Hole Thermodynamics}\label{class}

The previous two sections allow one to classify proofs based on how large and rapidly changing the perturbations to the black hole are permitted to be.  The next four sections offer a different classification based on the features of quantum mechanics which are included.

Consider first the regime in which any change in $S_{out}$ is much smaller than the changes in $S_{BH}$.  This means that quantum effects such as Hawking radiation are unimportant, leaving classical GR coupled to matter satisfying the null energy condition.  In this case the GSL reduces to the classical Second Law, which states that the area of the event horizon is nondecreasing.

In what situations is this approximation justified?  Suppose the black hole exchanges a small amount of Killing energy with a system outside the black hole.  The marginal entropy gain or loss in the systems is proportional to their inverse temperature.  So $\Delta S_{out}$ is negligible compared to $\Delta S$ whenever the Killing temperature of the external system is much larger than the temperature of the black hole.

In this regime, Hawking's area increase theorem \cite{hawking71} states that the area of all black hole event horizons increases with time.  This theorem requires an assumption related to cosmic censorship; the simplest assumption is that there are no singularities on the horizon.  Using this assumption I now give a rough sketch of the proof below:

Each horizon generator carries an infinitesimal amount of horizon area.  The change in this area over time is given by the Raychaudhuri equation:
\begin{equation}\label{Ray}
-\frac{d\theta}{d\lambda} = \frac{1}{2}\theta^2 
+ \sigma_{ab}\sigma^{ab}
+ 8\pi G T_{ab}k^ak^b,
\end{equation}
where $\theta = (1/A)({dA}/{d\lambda})$ is the expansion parameter, $\sigma$ is the shear tensor, and $k^a$ is a null vector on the horizon of unit affine length.\footnote{I.e. $\lambda_{;a} k^a = 1$ on the horizon generator.}  Since the right hand side of this equation is always positive by the null energy condition, a horizon generator with negative expansion is ``trapped'' and must terminate in the future at a finite value of the affine parameter.  It cannot terminate on a singularity because by assumption there are no singularities on the horizon.  Nor can it leave the horizon because it is impossible for generators to leave a future horizon.  Consequently, since all horizon generators have nondecreasing area and any new generators appearing on the horizon only add even more area, the area cannot decrease.  Consult Ref. \cite{WALDBOOK} for the full details of the area increase theorem.

This may be regarded as the first proof of the GSL, limited to the classical regime in which $S_{out}$ is negligible compared to $S_{BH} = A/4$.

\subsubsection{The Hydrodynamic Approximation}\label{hydro}

In quantum field theory (QFT) the entropy cannot be treated as a classical 4-vector, because it is not fully localizable.  Instead the entropy in quantum mechanics is subadditive, i.e. the entropy of a whole system can be less than the sum of the entropy of its parts \cite{wehrl78}.  Additionally, the entropy in a region with sharp boundaries is dominated by the divergent entanglement entropy of fields close to the boundary.  Some renormalization scheme is necessary to obtain a finite entropy.  In section \ref{FMW}, I argue that this can sometimes lead to superadditivity, in which the whole has more entropy than the parts.

However, in some situations the entropy is approximately localizable.  In this hydrodynamic approximation, the entropy and energy are described by classical currents $s^a$ and $T^{ab}$.  This is the setting for Wald \cite{wald94} (section \ref{atm}), and the proofs via Bousso's covariant entropy bound \cite{FMW00}\cite{BFM03} (section \ref{BousB}).

Unfortunately, I have not been able to find any regime in which this approximation is justified except when classical black hole thermodynamics is also valid.  This suggests that proofs using the hydrodynamic approximation are redundant, because they never apply except when classical black hole thermodynamics also applies.

To see the difficulty, consider blackbody radiation at local temperature $T$.  Quanta can only be considered well-localized at distance scales much larger than their average wavelength, which is inversely proportional to the local temperature $T$.  So a reasonable first guess would be that the hydrodynamic approximation is justified when the local thermodynamic potentials change significantly only over distance scales much larger than the inverse temperature.  But this condition does not seem to be satisfied by the thermal atmosphere near an event horizon, because its local inverse temperature is proportional to the proper distance from the horizon's bifurcation surface.  Since the thermal atmosphere cannot be accurately described by the hydrodynamic regime, it would appear that in the hydrodynamic regime can only apply to situations in which the thermal atmosphere can be neglected.  The only situation I know of like this is when the infalling matter has Killing temperature much larger than the temperature of the black hole---but then classical black hole thermodynamics also applies (cf. section \ref{class}), making the hydrodynamic regime redundant.

So further work should be done to explore when the hydrodynamic regime is really justified, in order to see exactly what new information the hydrodynamic proofs add beyond what was already given by the area increase theorem.

\subsubsection{The Semiclassical Regime}\label{SC}

Neither the classical nor hydrodynamic limits permit one to consider fully quantum mechanical states of matter using the techniques of QFT.  This deficiency is remedied by the semiclassical gravity approximation \cite{bardeen81}.  In this approximation the metric is treated as classical but it is coupled self-consistently to the expectation value of the renormalized stress-energy tensor via the semiclassical Einstein equation:
\begin{equation}\label{semiEin}
G_{ab} = 8\pi G \langle T_{ab} \rangle.
\end{equation}
Thus one neglects the gravitational effect of fluctuations in the stress energy tensor.  In the Feynman picture, this involves ignoring diagrams with graviton loops even while taking matter loops into account.

This approximation may be justified either in the large $N$ limit or in the quasi-stationary limit.  In the large $N$ limit, the contributions of each field to the expectation value of the stress-energy contribute coherently, and is therefore of order $N$ times the contribution of a single field.  On the other hand, the fluctuations of each field contribute incoherently and therefore are of order $\sqrt{N}$ times the fluctuations due to a single field.  So the matter fields can be arranged to have a large effect on the metric even while their fluctuations are negligible.  This permits exploration of the semiclassical but not quasi-stationary regime.

A difficulty arises, however, due to radiative corrections.  These can create higher-derivative terms in the gravitational action, leading to pathological extra degrees of freedom whose energy is unbounded below.  If the perturbation due to gravity is small, these extra degrees of freedom can be disposed of using perturbative constraints \cite{simon90}, but if the perturbation is large this method does not work.  Fortunately, there exist two-dimensional gravitational models without this problem.  This permitted Fiola et al. \cite{fiola94} to create a proof of the GSL set in the non-quasi-stationary regime using the RST model (section \ref{2D}).

The second situation in which the semiclassical approximation may be justified is in the quasi-stationary regime, in which the effect of the matter fields is a small perturbation to the metric.  One begins by specifying a classical background manifold (possibly sourced by some classical ``background'' stress-energy tensor) and then specifying a QFT state on this background manifold.  Because the perturbation to the metric is small in the quasi-stationary approximation, it is permissible to calculate the properties of this QFT state using the background metric instead of the perturbed metric.  In the case of quantum fields whose wavelength is of the order of a large black hole's radius $R \gg l_{P}$, the stress energy goes as $\langle T_{ab} \rangle \sim \hbar R^{-4}$, and the gravitational effects of the stress-energy on the metric are of order $\hbar G = l_{P}^2$ (the Planck length squared), which is small compared to $R^2$.  Gravitational perturbations are thus negligible except when they affect the Bekenstein-Hawking term $S_{BH}$.  Because $S_{BH}$ has an $l_{P}^2$ in its denominator (Eq. (\ref{Gen})), these $\mathcal{O}(l_{P}^2)$ perturbations of the geometry can produce an $\mathcal{O}(1)$ shift in the value of the generalized entropy.

One might worry that since the fluctuations in the stress-energy can be of the same order as the expected stress-energy, it is incorrect to treat the spacetime geometry as taking a definite value, invalidating Eq. (\ref{semiEin}).  However, this limitation is irrelevant for semiclassical proofs of the GSL if, as suggested by Ref. \cite{SS99}, $S_{BH}$ is taken as proportional to the \emph{expectation value} of the area (cf. section \ref{borg}).  Then all one needs is the expectation value of the first order change in the geometry, allowing Eq. (\ref{semiEin}) to be replaced with the expectation value of the linearized Einstein equation:
\begin{equation}\label{expEin}
\langle G_{ab}^{1} \rangle = 8\pi G \langle T_{ab}^{1} \rangle .
\end{equation}
This version of the semiclassical approximation still requires any fluctuations in the quantum fields to be small enough to neglect nonlinearities in the Einstein equation, but it does not require the fluctuations in the energy to be small compared to the average energy.

Since the gravitational field contains independent degrees of freedom, Eq. (\ref{semiEin}) is insufficient to completely determine the first order perturbation to the metric caused by the first order component of the stress-energy tensor.  In general this ambiguity must be resolved by an appropriate choice of boundary conditions, but fortunately proofs of the GSL may ignore this subtlety.  Why?  Because the only feature of the first order change in the geometry which must be considered to calculate the generalized entropy is the area, and the change in the area is given by the expansion parameter $\theta$.  Now $\theta$ can be calculated using the linearization of the Raychaudhuri equation (\ref{Ray}) about the background spacetime:
\begin{equation}
-\frac{d\theta^{0}}{d\lambda} = \theta^{0}\theta^{1}
+ 2\sigma_{ab}^{0}\sigma^{ab\,1}
+ 8\pi G\,T_{ab}^{1}k^ak^b.
\end{equation}
Imposing the event horizon final boundary condition $\theta |_{\lambda = \infty} = 0$, one can solve for $\theta^{1}$:
\begin{equation}\label{linearRay}
\theta^{1}(\lambda) = 
8\pi G \int^{\infty}_{\lambda} \!\! d\lambda^{\prime}
\,T_{ab}^{1}k^ak^b + 2\sigma_{ab}^{0}\sigma^{ab\,1}
\,\exp[\int^{\lambda^{\prime}}_{\lambda}\!\! \theta^{0} d\lambda^{\prime\prime}], \footnote{The effect of quantized gravitational wave excitations would be described using a fractional order term $\sigma_{ab} {}^{1/2}\sigma^{ab\, 1/2}$ in place of the $8\pi G T_{ab} k^a k^b$ term, both in this equation and below.}
\end{equation}  Therefore $\theta^{1}$ is a function of the source $T_{ab}^{1}$ alone iff the background shear tensor $\sigma_{ab}^{0}$ vanishes.

In the quasi-stationary case, the background value of $\sigma_{ab}^{0}$ does vanish, as well as $\theta^{0}$ and $T_{ab}^{0} k^a k^b$, and Eq. (\ref{linearRay}) becomes:
\begin{equation}\label{response}
\theta(\lambda) = 8\pi G \int^{\infty}_{\lambda} \!\! T_{ab} k^a k^b d\lambda^{\prime}.
\end{equation}
This equation can be used to determine the change in $\Delta S_{BH}$ from one time to another in the quasi-stationary regime.\footnote{As a bonus, if the GSL can be proven in the quasi-stationary case it can also be proven for small perturbations of classical \emph{non-stationary} black hole metrics.  By Hawking's area increase theorem (cf. section \ref{class}), if on any horizon generator, at some time, $\sigma_{ab}^{0}$ or $\theta^{0}$ is nonzero, then $\theta^{0}$ is positive prior to that time.  That implies that the GSL is automatically true up until that time, because the zeroth order area increase times $l_{P}^{-2}$ is of lower order in $l_{P}$ than any possible decrease in $S_{out}$ due to the dynamics of the quantum fields.}
\paragraph{The Entanglement Entropy Divergence}
Defining $\Delta S_{out}$ in the semiclassical regime is harder, because the entanglement entropy of any region with a sharp boundary diverges in QFT.  So in order to define a finite $S_{out}$, one must somehow subtract off this infinite entropy through a renormalization scheme.  Wald's proof in section \ref{atm}, because it remains in both the hydrodynamic and quasi-steady limits, can avoid this by only considering local changes to the entropy of the black hole's thermal atmosphere.  But proofs in the semiclassical regime must work harder: those by Zurek and Thorne \cite{ZT85} (section \ref{heur}) and Sorkin \cite{sorkin98} (section \ref{semi}) still require an explicit renormalization scheme.  Proofs using an S-matrix, such as Frolov and Page \cite{FP93} (section \ref{Smat}) or Mukohyama \cite{muko97}, evade this issue by only considering asymptotic quantum states.  However, this strategy can only be used to determine $S_{out}$ and $S_{BH}$ at the beginning and end of a perturbing process, making it unsuitable for proving the GSL for intermediate time periods except in the quasi-steady approximation, which permits one to find the intermediate values of the entropy by using a linear interpolation justified by Eq. (\ref{secundus}).

In order to analyze this divergence, it is necessary to impose some cutoff which regulates the infinite entanglement entropy, e.g. the t'Hooft ``brick wall'' cutoff \cite{thooft85}, in which the horizon is replaced with a reflecting boundary a proper distance $\delta$ from the bifurcation surface of a stationary black hole.  In four dimensions, the divergent part of the entropy is typically found to be something like:
\begin{equation}
S_{div} = kN\frac{A}{\delta^2} + \mathcal{O}(\ln\,\delta),
\end{equation}
where $N$ is the number of particle species evident at the cutoff scale $\delta$.\footnote{But see Ref. \cite{JP07} for a cutoff imposed in a freely falling frame which gives a different result.}

In order to define the GSL semiclassically, there should be some physically well-motivated renormalization procedure which makes changes in the generalized entropy finite.  This could be done by also making $S_{BH}$ diverge with the cutoff $\delta$ in an equal and opposite way from $S_{out}$, so that their sum is finite in the limit that $\delta$ becomes small (though still much larger than the Planck length, so as to remain in the semiclassical regime).  This dependence of $S_{BH}$ on $\delta$ is due to the renormalization of the gravitational coupling constants \cite{jacobson94}.  The RG flow of $G$ would absorb the divergences in the area term, while the RG flow of higher-order curvature couplings would cancel out the subleading divergences.\footnote{The modification of $S_{BH}$ induced by these terms may be calculated using the Noether charge method \cite{wald93}.  Since the identical changes to $S_{BH}$ also appear in the First Law (\ref{FIRST}) \cite{JKM95}, the basic structure of the semiclassical proofs presented here should be unaffected by these extra terms.}  Physically speaking, the idea is that some or all of the entropy attributed to the $S_{BH}$ term at long distance scales is actually revealed at short distance scales to be part of the entanglement entropy $S_{out}$.  It is thus natural that whatever is added to the latter term must be removed from the former term in order to avoid double counting the entropy.

If this interpretation is correct, the flow in the coupling constants needed to make the entanglement entropy finite should be the same as the ordinary RG flow needed to cancel the divergences of Feynman graphs.  Various one-loop calculations mostly support this correspondence, with a few anomalies \cite{FS94}.  However, the cutoffs in Ref. \cite{FS94} rely on a thermal exterior state on a stationary black hole in order to identify which state in the regulated theory corresponds to the thermal Hartle-Hawking state.  To apply these ideas to a proof of the GSL, one would need to find a more general regulator.

\subsubsection{Full Quantum Gravity}\label{QG}

Clearly the best proof of the GSL would be one valid in full quantum gravity.  Such a proof should reveal whether black hole thermodynamics is a substantive constraint on theories of quantum gravity or whether it is a generic feature of sufficiently ``good'' theories.  The other proofs would then be seen as special cases of this one.

However, no such proof can be made rigorous apart from a specific theory of quantum gravity, or at least a set of axioms describing a class of theories.  Since no fully satisfactory background free theory of quantum gravity exists, such proofs are very speculative.\footnote{The proposed duality between string theory on Anti-deSitter and certain Conformal Field Theories \cite{AdS/CFT} does not define a fully background free bulk theory, since it is limited to states which are asymptotically AdS.  Nevertheless it certainly describes a broad class of states in which there are black holes, so a proof of the GSL from the AdS/CFT duality would be highly significant.  See below for a sketch of how one might prove the GSL from this duality.}  In fact only one has been attempted, that of Sorkin \cite{sorkin86} (section \ref{full}).

Full quantum gravity must be able to describe Planck sized black holes, which have no separation of scale between quantum and gravitational effects.  Quantum fluctuations being large, the formalism must be capable of dealing with rapidly changing black holes, as well as quantum superpositions of any number of black holes---including none at all.  Even to formulate the meaning of the GSL in this context will be a great achievement.

If the full theory of quantum gravity cuts off the entanglement entropy at a particular distance of order $\delta = \sqrt{N}$ in Planck units, then the entire entropy of the black hole might be accounted for with the $S_{out}$ term alone \cite{sorkin83}\cite{jacobson94}.  This is the viewpoint taken by Sorkin's proof.  A single term is more parsimonious than a strange sum of two very different contributions.  It also justifies the renormalization of $S_{BH}$ described in section \ref{SC}, as the reflection of an arbitrary cutoff-dependent division of a conceptually single quantity into two component terms.  But it is difficult to reconcile a finite cutoff with the property of Lorentz symmetry \cite{jacobson00}, which is necessary for the GSL to hold (at least generically) \cite{EFJW07}.

It is believed by many researchers that the evolution and evaporation of a black hole is somehow described by a unitary S-matrix when full quantum gravity is taken into account \cite{preskill93}.  However, the loss of information in no way contradicts the laws of quantum mechanics, since it quite possible to describe quantum mechanical systems that leak out information (the positive trace-preserving linear maps of section \ref{full} give one possible way).  Every one of the proofs reviewed here permits information to be lost.  The proposal of unitary time evolution would imply that the semiclassical regime gives inaccurate results in a regime in which it might be expected to be valid.  It also appears to be radically nonlocal unless its principles can also be also be extended to arbitrary Rindler horizons, which cannot be locally distinguished from black hole horizons.\footnote{A referee suggests an argument that this unitary hypothesis is also incompatible with the GSL.  Suppose a black hole of area $A$ forms from the collapse of matter in a pure state, and $S_{out} > -A/4$, so that the generalized entropy increases.  Then if the black hole completely evaporates, the state must be pure by virtue of the unitary S-matrix, and the generalized entropy becomes zero again.

One possible response is that the argument that the black hole entropy initially increases is based on semiclassical principles, while the argument that the state is pure at the end is based on full quantum gravity principles.  If the semiclassical picture is obtained from the full theory by some sort of coarse-graining procedure, then changing regimes in the middle of the argument may be invalid.  One could make an analogy to the ordinary thermodynamics of a box of gas which begins in a pure state at time $t_1$.  From a coarse-grained perspective, the entropy in the box increases with time from $t_1$ to $t_2$, but from the fine-grained perspective it remains pure even at a later time $t_3$.  This ``decrease'' of entropy from $t_2$ to $t_3$ is an artifact of changing perspectives and should not be deemed a violation of the OSL.}

Nevertheless, suppose one were to postulate unitary time evolution on slices which are complete outside the event horizon (this ``outside unitarity'' assumption is stronger than simply requiring the S-matrix to be unitary).  Further assume that the entire generalized entropy of the black hole really comes from the $S_{out}$ term alone.  Under these assumptions the GSL could be proven in exact analogy to the OSL.  Trivially, the fine-grained Gibbs entropy neither goes up nor down under unitary evolution.  However, to recover the entropy increase found in the semiclassical limit one would then have to impose some additional form of coarse graining, aside from the horizon (since under the unitary hypothesis the horizon conceals no information).  The challenge to those who believe in unitary outside evolution is to define this coarse grained entropy, and to show that it reduces to the generalized entropy in the semiclassical limit.

A similar kind of proof might be possible in the case of AdS/CFT.  Even if the outside unitarity assumed by the preceding paragraph is too strong to be true, the fact that the conformal field theory has unitary time evolution means that one might try to prove the GSL in the bulk from the OSL on the boundary.  Assuming that the duality is exact, one would need to identify a coarse-grained entropy on the boundary theory and show that this coarse-grained entropy both increases and is identical to the generalized entropy in the bulk theory.

\subsection{Are the Entropy Bounds necessary for the GSL?}\label{bounds}

It is often asserted that the GSL limits the amount of entropy capable of being stored in a region.  The most important proposals for the purposes of this review are Bousso's covariant entropy bound \cite{bousso99} and the Bekenstein bound \cite{bek81}.

Bousso's bound states:  Suppose one takes any spatial 2-surface $B$ with area $A$, and shoots out from it a normal lightsurface $L$ in any of the four possible directions.  Then as long as $L$ is initially contracting everywhere, the entropy $S$ passing through $L$ is bounded by
\begin{equation}
S \le \frac{kA}{4G\hbar}.
\end{equation}

To support the Bousso bound, one might argue that if $B$ is a cross-section of a black hole event horizon, and $L$ the horizon prior to $B$, a violation of the Bousso bound would mean that more entropy would fall into the black hole than is accounted for by its current entropy.  Alternatively one might argue that if $L$ completely encloses the past or future of an ordinary region of spacetime, and yet more entropy is found inside than permitted by the Bousso bound, adding more energy to the region would make it collapse into a black hole of the same area and thus the GSL would be violated.  However, neither of these arguments is very convincing.  Suppose that the Bousso bound is violated due to a large number of particle species, or due to some hyper-entropic object carrying a large number of degrees of freedom in a small space.  Then these objects ought to feature prominently in the black hole's thermal atmosphere, leading to additional large contributions to $S_{out}$.  These contributions can salvage the GSL in such cases \cite{MS03}.

Similarly, the Bekenstein bound states \cite{bek81} that in an isolated and weakly self-gravitating region of characteristic length $R$ and energy $E$, the entropy $S$ satisfies
\begin{equation}\label{Bek}
S \le \frac{2\pi k}{\hbar}RE.
\end{equation}
(Bekenstein took the characteristic length $R$ to be the widest dimension of the system, but it has also been argued that the bound should refer to the thinest dimension \cite{bousso02}.)  The Bekenstein bound's motivation is similar to that of the Bousso bound, but instead of collapsing the entire system into a black hole, one adds it to a preexisting black hole.  One possibility is that the system violating the bound is placed in a box and then slowly lowered into the black hole.  By means of the First Law (\ref{FIRST}), one then appears to obtain a violation of the GSL \cite{bek81} (cf. section \ref{horizon} for a more detailed example of this argument).  However, Unruh and Wald \cite{UW82} showed that the thermal atmosphere of a black hole acts on the box with a buoyancy force.  This prevents the box from being lowered closer to the horizon than its ``floating point'' without expending work, and is sufficient to save the GSL from being violated by the box.

Alternatively the system may be released from far away and allowed to fall into the black hole as in Ref. \cite{bek01}, which derives Eq. (\ref{Bek}) though with a somewhat larger numerical coefficient.  However, like the argument above for the Bousso bound, this calculation does not take into account the fact that if hyper-entropic objects exist, they will also be Hawking radiated by the black hole, again plausibly saving the GSL \cite{MS03}.\footnote{Bekenstein's rejoinder \cite{bek04} that such hyper-entropic objects would take too long to form is unpersuasive because the thermal atmosphere originates from extremely high frequency degrees of freedom in the local vacuum state.  According to the Unruh effect, such degrees of freedom are already in a perfect thermal state in every QFT with local Lorentz symmetry \cite{BWUW}, making their timescale of formation and dissolution irrelevant.  The objection can be sustained only if there is a breakdown of perfect Unruh thermality in quantum gravity, but such an effect would probably doom the GSL regardless of whether the bounds are satisfied.  Also, none of the proofs in sections \ref{osl}, \ref{Smat}, \ref{full}, \ref{semi}, or \ref{2D} assume anything similar to either bound, which suggests that neither bound is necessary for the GSL to hold.}

Note that Newton's constant $G$ is nowhere to be found in Eq. (\ref{Bek}).  The bound is motivated by gravitational physics and yet would constrain physics even in the QFT regime, by ruling out more than an order unity (though large) number of particle species \cite{page82}.  Bekenstein claims that his bound is saved even in the case of large number of species because of the Casimir energy of the large number of particle species \cite{bek94b}.  Responses to this claim were given by Page \cite{page04}, and Marolf and Roiban \cite{MR04}.

Despite the fact that the GSL does not imply either of the bounds, the converse statement that the bounds imply the GSL appears to be close to true in certain limits.  The proofs of the GSL in section \ref{BousB} begin by formulating ang proving a strengthened version of the Bousso bound, which in turn implies the GSL in the hydrodynamic approximation.  Since the Bousso bound as presently formulated does not hold in every situation \cite{lowe99}, these proofs must work from more restrictive assumptions than those necessary for the GSL.  In one of these proofs, the assumption added is similar to the Bekenstein bound (section \ref{FMW}).

\section{Proofs applying the OSL to the Thermal Atmosphere}\label{osl}

\subsection{Proof by Analogy to an Ordinary Blackbody System}\label{heur}

Zurek and Thorne (ZT) provided one of the first proofs of the GSL \cite{ZT85}.  Though the details are not as clear as in some later proofs, their argument was a major influence on many of the later proofs.  ZT begin by assuming that the entropy of a black hole is entirely due to the entanglement entropy in the thermal atmosphere.  This assumption is bolstered by a quasi-steady calculation of the total number of ways to build up a black hole by injecting quanta into the modes of the thermal atmosphere.  The resulting entropy equals the Bekenstein-Hawking entropy.

ZT proceed to write:
\begin{quote}\small
The above analysis provides, as a side product, a proof of the generalized second law of thermodynamics---that in any process involving the interaction of a black hole with the external universe, the sum of the black hole's entropy and the universe's entropy cannot decrease.  The proof: Since the hole's atmosphere plays the role of a thermal bath which exchanges particles with the universe, and since (when one used energy at infinity $\epsilon$ and Hawking temperature $T_{H}$ instead of locally measured energy E and temperature T) the change in the hole's entropy is precisely that associated with a standard thermal bath, the generalized second law is merely a special case of the ordinary second law.  (\cite{ZT85} p. 2174)
\end{quote}
Thorne, Zurek, and Price (TZP) have a more developed version of this argument in a book on the membrane paradigm \cite{TZP86}.  This paradigm is an elaborate mathematical analogy between a quasi-steady black hole and a viscous 2-dimensional fluid membrane located an infinitesimal distance outside of the black hole horizon, and coupled to the fields outside the membrane by various boundary conditions.  So long as one only cares about what happens outside of the black hole, the evolution of the exterior system coupled to the membrane is equivalent to the coupling to the black hole interior.  In this framework, TZP argue that:
\begin{quote}\small
From the discussion and equations in the last subsection it should be clear that whenever a slowly evaporating black hole interacts with the surrounding universe, its statistical properties [...] are exactly like those of an elementary, nongravitating but rotating thermal reservoir.  Compare, e.g. the probability distributions for the number of quanta in each mode of the field in the perfectly thermalized limit [...] or the expressions for the entropy changes resulting from interaction with the external universe. [...] Since the standard derivations of the second law of thermodynamics are perfectly valid for arbitrary systems interacting with such an elementary reservoir, it is clear that they must be equally valid for arbitrary systems interacting with a slowly evolving black hole.  Thus \emph{the second law of thermodynamics is just a special case of the standard second law of thermodynamics.  In such a system the total entropy, including that of matter and fields contained outside of the hole's stretched horizons, can never decrease} [emphasis theirs].  (\cite{TZP86} p. 313)
\end{quote}
This verbal argument does not specify what ``standard derivation of the [ordinary] second law'' should be used as the basis for the proof.  TZP thus need the reader to supply some interpretation in order to turn the argument into a complete proof.  My attempt at interpretation now follows:

The entropy of the system is the sum of the elementary thermodynamic entropy of the ``elementary, nongravitating but rotating thermal reservoir'' (i.e. the membrane), and the system exterior to the membrane.  One may write this as
\begin{equation}\label{sep}
\Delta S = \Delta S_{BH} + \Delta S_{out},
\end{equation}
where $S_{BH}$ represents the entropy of the membrane, and $S_{out}$ represents the entropy outside the membrane.  Moving the membrane closer to the horizon ought to renormalize the black hole entropy as described in section \ref{SC}, by decreasing the value of $S_{BH}$ and increasing the value of $S_{out}$ to compensate (assuming for the moment that $S_{BH}$ and $S_{out}$ are finite and well defined).

In order to successfully correspond with the black hole system, one must also be able to identify $S_{BH}$ with the entropy stored in the layers of thermal atmosphere between the horizon and the membrane (call this the ``deep atmosphere''), so that the generalized entropy is the same in both systems---otherwise a proof that entropy increases for the membrane system will not carry over to the analogous black hole system.  When the membrane is far from the horizon, this ``deep atmosphere'' is the whole atmosphere, and should thus be equal to a quarter of the area of the horizon by virtue of the calculation in ZT \cite{ZT85}.

It can be calculated---at least for free fields and quasi-steady black holes---that the membrane absorbs everything that falls on it and emits only exact thermal radiation.  From this it follows that anything that falls into the deep atmosphere can be treated as though it were exactly thermalized.

Armed with the above results, the correspondence between the black hole system and the membrane system can be shown.  In the quasi-steady limit, both the membrane and the deep atmosphere obey the Clausius relation (the former because of the First Law of black hole thermodynamics, and the latter because anything that falls into the deep atmosphere can be treated as if it thermalizes):
\begin{equation}
\Delta E = T \Delta S.
\end{equation}
Therefore, whenever matter falls into the deep atmosphere, one replaces the state of the deep atmosphere with another in which the infalling energy is fully thermalized amongst all the degrees of freedom in the deep atmosphere.  This can only increase the entropy.  This thermalized deep atmosphere then behaves equivalently to the membrane system, for which a second law holds.  Since both of these processes increase the entropy, the GSL always holds.

As far as I can tell, this argument is equivalent to the thin shell argument presented by Wald \cite{GW01}\cite{wald01}, with the ``thin shell'' being another name for the ``elementary thermal reservoir''.

\paragraph{Limitations}

What can go wrong here?  The most serious problem is the absence of a regularization scheme needed to make $S_{BH}$ and $S_{out}$ finite.  Both the horizon and the membrane are sharp boundaries, and are therefore each associated with infinite entanglement entropy.  The horizon entanglement makes $S_{BH}$ diverge, and the membrane entanglement makes both $S_{BH}$ and $S_{out}$ diverge.  The entanglement across the membrane makes the total entropy subadditive, thus invalidating the separation into two terms of Eq. (\ref{sep}), since the entropy cannot in fact be fully localized (cf. section \ref{hydro}).  Therefore a justification of the correspondence between the black hole and the membrane picture requires serious work before it can be considered well-defined.

As an alternative interpretation of TZP's argument, one might admit that the black hole system stands in need of regularization, but suggest that the membrane paradigm is itself the regularization scheme needed to render the black hole entropy finite.  This interpretation would view the correspondence between the black hole and the membrane not as a mathematical identity between two distinct well-defined systems, but rather as a formal identity between the unregulated and ill-defined entropy of the black hole system, and a regulated well-defined membrane system.  Replacing the deep atmosphere with the membrane would itself be the way to regulate the generalized entropy.

The trouble with this interpretation is that it is not clear that the entropy and dynamics of the membrane are really completely mathematically well-defined.  Although the black hole does seem to behave like a membrane for the purposes of the several calculations listed by TZP above, in order to be completely well-defined semiclassically, one would have to be able to fully specify the interactions between the membrane and the dynamics in all QFT states.  The membrane satisfies an idealized blackbody condition: it absorbs everything that impinges upon it while emitting exact thermal radiation.  Unlike the usual (e.g. reflecting) boundary conditions, this boundary condition permits the loss of information, meaning that the fields coupled to the boundary condition do not evolve according to unitary dynamics coming from a Hamiltonian.  I do not know how one would quantize such a field theory, nor am I aware of any work on this subject.

\subsection{Proof by Perturbing the Thermal Atmosphere}\label{atm}

Rather than create an analogue membrane or shell system like the proofs in the previous section, Wald \cite{wald94} obtains his proof by describing changes in the thermal atmosphere  In order to sidestep the problems with entropy localization, he describes this atmosphere using the hydrodynamic regime, in which the entropy outside of the black hole is can be approximated by a classical current---i.e. it is fully localizable.  Then he considers infalling matter, which must be in the form of a small quasi-steady\footnote{In Ref. \cite{wald94}, Wald considers arbitrary small quasi-stationary perturbations, but this is only enough to get entropy increase over the course of the entire process (cf. section \ref{quasi}).} perturbation of this thermal atmosphere to obtain the GSL.  By bounding the amount by which this perturbation can increase the atmosphere using the Clausius relation from ordinary thermodynamics, Wald is able to limit the change in $S_{out}$ based on the amount of energy flowing into the black hole.  The amount of energy flow also determines the change in $S_{BH}$ by means of the First Law of black hole thermodynamics, resulting in a proof of the GSL.

In the Hartle-Hawking state, a stationary black hole is surrounded by a thermal atmosphere.  Locally this radiation looks just like blackbody radiation.  Therefore fiducial observers co-rotating just outside the horizon will observe an energy density profile of the form
\begin{equation}
e = T_{ab}\,\xi^a \xi^b / \xi^2,
\end{equation}
where $\xi$ is the Killing field which generates the horizon, and $T_{ab}$ is the expected stress-energy difference between the Hartle-Hawking state and the vacuum with respect to the Killing flow (i.e the Boulware state).  These fiducial observers should also see an entropy density
\begin{equation}
s = S_a\,\xi^a / \xi,
\end{equation}
where $S_a$ is the entropy current associated with the thermal radiation observed by fiduciary observers.

In the Hartle-Hawking state, the outgoing Hawking radiation is exactly balanced by incoming thermal radiation.  Wald now modifies this incoming state by a small perturbation.\footnote{This will result in a slightly different spacetime due to gravitational interactions.  To compare the results of the original and final spacetimes, Wald uses diffeomorphism symmetry to identify points in such a way that the Killing field $\xi$ of the unperturbed spacetime has the same norm at identified spacetime points.  However, because the gravitational effects are a small perturbation, it is acceptable to consider the entire process as taking place on one background spacetime (cf. section \ref{SC}).  The only relevant gravitational effect is the infinitesimal change in the horizon area.}

The perturbation in the energy density is
\begin{equation}
\delta e = \delta [T_{ab}\,\xi^a \xi^b / \xi^2] = (\delta T_{ab}) \xi^a \xi^b / \xi^2,
\end{equation}
and similarly the perturbation in the entropy is
\begin{equation}
\delta s = \delta [S_{a}\,\xi^a / \xi] = (\delta S_{a}) \xi^a / \xi.
\end{equation}
Any ``small'' perturbation to a thermal state satisfies the Clausius relation:
\begin{equation}\label{Clausius}
\delta s \le \delta s_{th} = \delta e / T = 2\pi \xi \delta e / \kappa
\end{equation}
where $s_{th}$ is the entropy if the final state is still perfectly thermalized.  Taking the limit as the fiducial observers approach the horizon, and multiplying by $\xi$, Wald obtains
\begin{equation}\label{tofirst}
-(\delta S_a) \xi^a |_{horizon} \le 
{2\pi \over \kappa} (\delta T_{ab})\xi^a \xi^b |_{horizon}.
\end{equation}
Wald integrates both sides of this inequality over the horizon, including the null direction.  The left hand side becomes the total entropy falling through the surface as a result of the perturbing process, while the right hand side becomes the change in $A/4$ given by the First Law (\ref{first}) for all quasi-steady physical processes.

But by the OSL, $S_{out}$ cannot be reduced by more than the entropy flowing into the black hole.  It follows that
\begin{equation}\label{GSLresult}
-\Delta S_{out} \le \Delta A/4,
\end{equation}
which is the GSL.

\paragraph{Limitations}

How ``small'' does the perturbation of the black hole have to be for this proof to apply?  The bottleneck is in the use of the Clausius relation on line (\ref{Clausius}): only for a first order increase in energy is it generally true that $\delta s_{th} = \delta e / T$, since to second order the temperature of the state changes.  Consequently, the proof as it was written appears to require the adiabatic regime, in which the atmosphere is only modified by a first order perturbation.  But for first order changes of the state, the Clausius relation $\delta s = e / T$ is actually an equality rather than an inequality, so that Eq. (\ref{GSLresult}) also becomes an equality:
\begin{equation}\label{asanequality}
-\Delta S_{out} = \Delta A/4.\footnote{By the argument in section \ref{adiabatic}, this result must hold for all adiabatic processes even if they are not quasi-steady.  This gives rise to an apparent violation of the GSL if one sends in an adiabatic pulse of energy with no support prior to an advanced time $t$.  Because of the teleological boundary condition, the horizon grows in anticipation of the energy which is to come, so it seems that initially $S_{BH}$ increases while $S_{out}$ remains the same.  But then by Eq. (\ref{asanequality}), the generalized entropy remains the same at the beginning and end of the process, which means that it must decrease at some later time to counterbalance its initial increase.  But that violates the GSL.  Presumably the solution is that any quantum state has long distance entanglements not taken into account in the hydrodynamic limit, which affect $S_{out}$ even before the advanced time $t$.}
\end{equation}
This would mean that the proof would have have a very limited range of applicability.

However, it is possible to free this proof from the assumption that the perturbation be adiabatic.  This assumption justifies the Clausius relation (\ref{Clausius}), which bounds the entropy in the thermal atmosphere given a small change in its energy density.  Assuming that the energy density $\Delta e$ of the perturbation is large enough to meaningfully change the local temperature, Eq. (\ref{Clausius}) no longer applies.  Let $T(e)$ be the temperature of thermal equilibrium at an energy density $e$; then the change in entropy is given by an integral:
\begin{equation}\label{nonlinear}
\Delta s \le \int^{e + \Delta e}_e \!\frac{de^\prime}{T(e^{\prime})}.
\end{equation}
Since the heat capacity of blackbody radiation is positive (at least for weak interactions), adding a finite amount of energy density increases $T$ in the denominator and thus makes the constraint on $\Delta s$ even more stringent than that given in (\ref{Clausius}).  On the other hand, if energy is removed from the thermal atmosphere this decreases $T$ in the denominator, which because of the change in the sign of $e$, also leads to a more stringent constraint in $\Delta s$.  So as long as the thermal atmosphere has positive heat capacity, there is no need to consider adiabatic perturbations; quasi-steady perturbations are small enough.\footnote{As an alternative to this argument, in the limit that the fiducial observers approach the horizon, the change of temperature should become less and less important in all dimensions $d > 2$.  Neglecting factors of order unity, the heat capacity of blackbody radiation is \begin{equation}\label{cap}
C = VT^{d - 1},
\end{equation}
where $V$ is the volume and $T$ is the temperature defined with respect to the proper time of the local fiducial observer.  If the fiducial observer is at proper distance $x$ from the bifurcation surface, it sees a local temperature $T = 1/x$.  When a pulse of energy falls into the black hole at a fixed retarded time, a fiducial observer closer to the horizon will see this pulse in its own frame of reference as having energy proportional to the scaling factor $x^{-1}$, and volume proportional to $x$.  This energy pulse is viewed by the fiducial observer as raising the energy of a heat bath of equal volume whose total heat capacity $C$ therefore scales as $x^{2 - d}$.  Multiplying both sides of Eq. (\ref{nonlinear}) by the volume, and expanding the result out as a power series in the added energy $\Delta E$, one obtains
\begin{equation}
\Delta S \,\le\, 
\frac{\Delta E}{T_{0}} \,-\, \frac{(\Delta E)^2}{2CT_{0}^2} \,+\, \mathcal{O}(\Delta E^3),
\end{equation}
where $T_{0}$ is the temperature prior to the perturbation.  The first nonlinear correction term now scales as $x^{d-2}$ since $T$ and $\Delta E$ scale together, leaving only the scaling of the heat capacity in the denominator.  The higher order terms will be even more suppressed.  This shows that for $d > 2$, any dose of energy falling into the black hole is ``small'' enough to render Eq. (\ref{Clausius}) valid.  In the case of interacting fields, there will be corrections to Eq. (\ref{cap}).  However, the only property of  Eq. (\ref{cap}) needed is that the heat capacity of blackbody radiation increases without limit as the temperature increases.  It is difficult to imagine any sensible QFT with $d > 2$ violating this assumption, since this would require that the heat capacity in the interacting theory differ from the heat capactity in the free theory by an arbitrarily large factor in the high energy limit.}

Therefore, there is good reason to believe that Wald's proof can be relieved of the need to assume adiabaticity in most settings.  But the proof still relies crucially on the hydrodynamic assumption that entropy can be fully localized, which is not even fully true in classical mechanics and which goes very wrong in QFT.  The hydrodynamic approximation is likely to be especially inaccurate when applied to the thermal atmosphere of a black hole (cf. section \ref{hydro}).  It is difficult to see how to modify the proof in a way that gets around this assumption, given its heavy use of the concept of local thermal equilibrium.

\section{Proof using the S-Matrix}\label{Smat}

\begin{figure}[ht]
\centering
\includegraphics{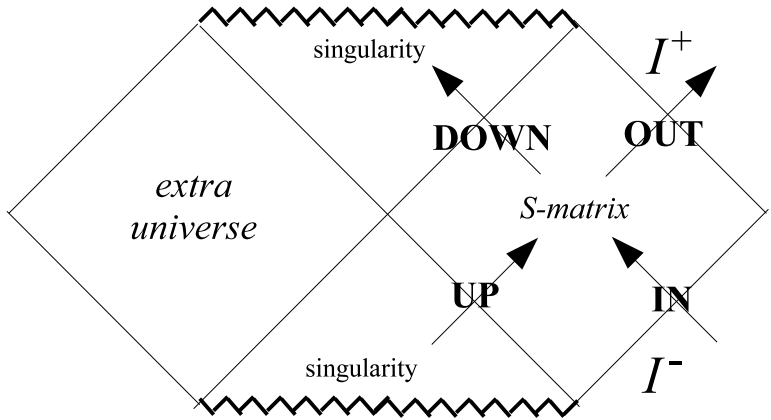}
\caption{The Penrose diagram of an eternal black hole.  The S-matrix is used to evolve the UP and IN modes into the DOWN and OUT modes.  In the case of the black hole which forms from collapse, the white hole horizon is replaced by the collapsing star and the UP modes are populated by the Hawking effect.}\label{eternal}
\end{figure}

Frolov and Page (FP) \cite{FP93}, inspired by the arguments of Zurek, Thorne, and Price \cite{ZT85}\cite{TZP86} (section \ref{heur}), provided a straightforward and explicit proof of the GSL for semiclassical, quasi-steady black holes.  In the quasi-steady limit, any processes taking place over a finite period of Killing time may be described using a stationary black hole metric.  These interactions can be described by a unitary S-matrix relating the asymptotically past density matrix $\rho_{past}$ to the asymptotically future $\rho_{future}$.  The information in $\rho_{past}$ consists of the infalling ``IN'' modes and the ``UP'' modes populated either by the white hole horizon (in the eternal case), or by the Hawking effect (if the black hole formed from collapse).  Similarly, $\rho_{future}$ specifies both the ``DOWN'' modes falling through the black hole horizon and the ``OUT'' modes radiated to infinity  (see Figure \ref{eternal}).
The advantage of the S-matrix formulation is that it allows one to bystep the divergence of $S_{out}$ at the horizon, by only considering the entropy when it is infinitely distant from the black hole.\footnote{Admittedly, the changes in the entropy and energy of the outside matter are still technically infinite, since the S-matrix is only defined in the limit of infinite time, and the quasi-steady assumption approximates the entropy and energy flux into the black hole as being constant with time.  However, this divergence can be removed by simply dividing all such quantities below by the total time elapsed.}

So far everything is time reversal symmetric.  To get the GSL, FP also need to assume that: i) the UP state consists of radiation at the Hawking temperature, and ii) the UP state is uncorrelated with the IN state.

In the eternal case these assumptions both hold if one begins with the Hartle-Hawking state and arbitrarily adjusts the IN state without changing the UP state.

In the collapsing case the assumptions are reasonable in the semiclassical picture, in which the UP mode thermal radiation can be traced back to Unruh radiation at the formation of the event horizon.  Since the black hole must eventually become quasi-steady for this proof to hold, this radiation traces back to exponentially high frequencies and so can be expected to be essentially in the vacuum state regardless of the matter state used to form the black hole \cite{FP93}.  Therefore there is good reason to believe that the collapsing case can be well approximated by uncorrelated UP and IN modes.

Since the S-matrix is unitary, FP now invoke the OSL to show that
\begin{equation}\label{FPosl}
S_{U} + S_{I} = S_{past} = S_{future} \le S_{D} + S_{O},
\end{equation}
using the lack of correlation between UP and IN, and also the subadditivity of entropy for DOWN and OUT.

FP now apply the First Law of black hole thermodynamics (\ref{first}) to the temperature T and energy E observed by a fiducial observer just outside and co-rotating with the horizon:
\begin{equation}
dS_{BH} = T^{-1}dE.
\end{equation}
In the semiclassical, quasi-steady approximation, the change in energy of the black hole is equal to the expectation value $\langle E_{D} - E_{U} \rangle$, while $T$ remains constant, so that
\begin{equation}\label{BHchange}
\Delta S_{BH} = T^{-1} \langle E_{D} - E_{U} \rangle.
\end{equation}

Combining the change in the black hole entropy given by (\ref{BHchange})
with the change of matter entropy given by (\ref{FPosl}), FP find that
\begin{eqnarray}
\Delta S = \Delta S_{BH} + \Delta S_{out} = T^{-1}\langle E_{D} - E_{U} \rangle + S_{O} - S_{I}
   \\*	
   \ge (S_{U} - T^{-1}\langle E_{U} \rangle) 
     - (S_{D} - T^{-1}\langle E_{D} \rangle).\phantom{MMM}
\end{eqnarray}
The quantity $S - T^{-1} \langle E \rangle $ is equal to minus the free energy divided by the temperature.  This quantity is maximized in a given system when it is at the thermal state of temperature T, in which case its value is equal to $\ln \: Z$, Z being the partition function.  Thus, as long as the partition functions are equal for the UP and DOWN systems, 
$\Delta S \ge 0$.

Why should these systems have the same partition function?  FP suggest that this follows from CPT symmetry.  However, this argument is insufficient for the case of charged black holes, because the UP modes of a positively charged hole would be related by CPT to the DOWN modes of a \emph{negatively} charged black hole.  What is needed is a relation between the UP and DOWN modes of the same black hole.  This difficulty may be solved by appealing to the property that the partition function is multiplicative for independent subsystems, which implies that
\begin{equation}
\ln \:Z_{U} + \ln \:Z_{I} = \ln \:Z_{past} = \ln \:Z_{future} = \ln \:Z_{D} + \ln \:Z_{O},
\end{equation}
and thus to prove $Z_{U} = Z_{D}$ it is sufficient to show that $Z_{I} = Z_{O}$.  The latter may now be directly established by CPT since the black hole's charge should make no difference to the dynamics of these asymptotically distant modes.  However, perhaps it is better to avoid any reference to time-reversal symmetry and simply note that the possibility of providing unitary energy-conserving boundary conditions at spatial infinity relating the OUT and IN modes requires that their partition functions match.  Then the proof might be capable of extension to exotic CPT-violating theories.\footnote{However, such theories must also violate Lorentz invariance \cite{greenberg06}, which seems in general to lead to a failure of black hole thermodynamics due to UP modes no longer being thermal \cite{EFJW07}.}

\paragraph{Limitations}

Mukohyama has claimed that FP's proof applies only to the eternal black hole case, and fails when extended to collapsing black holes \cite{muko97}.  His reasoning is that when the black hole forms from collapse the information in the UP modes comes originally from incoming matter prior to the formation of the event horizon.  Therefore if the incoming matter at earlier times is entangled with incoming matter at later times, the UP and IN modes will be correlated.  This situation violates assumptions i) and ii) above, which are required for FP's proof.

This criticism does not seem to be relevant to FP's proof because it uses the quasi-steady limit.  Although the S-matrix is also defined using a very long time interval between the initial and final states, the period of time over which the black hole grows from collapsing matter must be far longer---or else FP could not have used the S-matrix elements defined on a stationary background in their proof.  In this limit all of the contaminated UP modes have plenty of time to either fall into the black hole or escape to infinity, before the beginning of the period analyzed by FP.  The UP modes that become relevant to the proof are in the extreme UV at the time of formation and are therefore unaffected by the particular state of the infalling matter.  Of course, any generalization to the collapsing case that went beyond the quasi-steady limit would have to deal with the issue Mukohyama raises, but on its own standards the proof applies equally to the eternal and collapsed cases.  (Cf. section \ref{comb} for discussion of Mukohyama's proposed extension \cite{muko97} of FP's proof to the collapsing, but still quasi-steady case.)

A more serious limitation is that this proof cannot be applied to a black hole system enclosed in a finite sized box.  Such a box would reflect OUT modes into IN modes which would generally lead to correlations between the UP and IN modes, violating assumption ii).  It would also make it impossible to regard IN as temporally prior to OUT, invalidating the commutation relationships implicit in the S-matrix picture.  For example, suppose a particle carrying a qubit of information falls in from the boundary, scatters off the black hole, bounces off the boundary and falls in a second time.  Describing this situation with the S-matrix above would lead to a duplication of quantum information, with the qubit appearing twice in the IN state.  In this context it is not natural to make a sharp division between IN, OUT, UP, and DOWN states; it makes more sense to look at the state as being defined on an achronal time slice and ask how it evolves to future slices.  This approach is used by the proofs in the next section.\footnote{Note that these difficulties do not apply to the boundary at ``infinity'' used in the partition function argument above, since in this case the box reflects radiation back on a timescale larger than the timescale for which the quasi-stationary S-matrix is well-defined.  Therefore it does not forbid the separation of UP and IN modes over the period of time needed for the proof.}

\section{Proofs from a Time Independent State}\label{special}
This kind of proof, due to Sorkin, begins by defining a special mixed state corresponding to the thermal state outside of the event horizon of the black hole.  Astonishingly, one can show that if this particular state evolves to itself, then there is a quantity which is nondecreasing under time evolution for all states.  If this nondecreasing quantity can be equated with the generalized entropy, this results in a proof of the GSL.

Sorkin created two different proofs using this method: one applying to the full quantum gravity regime \cite{sorkin86}, and the other to the semiclassical quasi-steady regime \cite{sorkin98}.  Unfortunately, neither proof appears to be sound as it stands.  The full quantum gravity proof has inconsistent assumptions, while the semiclassical proof has an unwarranted step.

Mukohyama also has a semiclassical quasi-steady proof \cite{muko97} combining this method with the S-matrix approach of section \ref{Smat}.  His proof and Sorkin's semiclassical proof both run into difficulty when applied to rotating black holes due to the absence of a well-defined Hartle-Hawking state for Kerr black holes (cf. section \ref{semi}).

\subsection{Full Quantum Gravity Version}\label{full}

The key feature of this proof \cite{sorkin98} is the use of a remarkable theorem:

Theorem 1: \emph{Given a quantum system with a finite dimensional Hilbert space, and a positive trace-preserving linear map on the space of density matrices, if the uniform probability state evolves to itself, then any state always evolves to a state with greater or equal entropy.}

(I have stated Theorem 1 as it is proven by Sorkin himself in Ref. \cite{sorkin98}.  However, it is a special case of a much more general result concerning the nonincrease of the ``relative entropy'' proven in Ref. \cite{lindblad75}.  In its most general form this result can be applied to arbitrary observable algebras.)

If one applies Theorem 1 to the system outside the horizon, a proof of the GSL requires only a few more steps.  First, one must argue that in the full quantum gravity regime, the generalized entropy is really given by just the $S_{out}$ term.  This would be true if the entropy associated with the area is entirely due to the entanglement entropy across the horizon.  If quantum gravity somehow cuts off the entanglement entropy at distances the order of the Planck length, and the effective number of propagating fields is of order unity, one obtains an entropy per area of the same order as the Bekenstein-Hawking entropy, lending credence to the idea that it is simply a form of entanglement entropy \cite{sorkin83}\cite{jacobson94}.

Second, one must show that the hypotheses of the theorem apply to the system outside the horizon, so that the outside entropy $S_{out}$ cannot decrease.  Sorkin needs additional assumptions to prove this result.  Before specifying a particular mathematically rigorous theory of full quantum gravity, it is impossible to know for sure that any of these assumptions are sound.  However, one may appeal to those features of QFT and GR which might plausibly apply to quantum gravity.  I have rephrased and reordered Sorkin's assumptions below, and also filled in some steps implicit in his argument:

\newcounter{comment}
\setcounter{comment}{0}

\begin{enumerate}

\item It makes sense to talk about the region of spacetime $\mathcal{R}(t)$ containing everything which is outside of the event horizon of a black hole at a given time $t$, and to assign this region an algebra of observables $\mathcal{A}(t)$.

\setcounter{comment}{\value{enumi}}
\end{enumerate}
\noindent
For example, in GR with Anti-deSitter boundary conditions, one may pick a time coordinate $T$ on the conformal boundary and then covariantly define the region as the union of the future of the $T=t$ locus on the boundary, with the region causally to the past of the boundary.\footnote{Sorkin's language in Ref. \cite{sorkin86} associates the observables with a spacelike slice going from the boundary of the spacetime to the horizon.  On the assumption that the observables are causal this is equivalent to the language I use here.}  In quantum gravity, there may be large quantum superpositions of spacetime geometry, so this ``region'' might have very different geometries in different branches of the superposition.  Due to quantum fluctuations there might even be no black hole or multiple black holes. Is it meaningful to assign a fixed algebra to such a wildly varying region?  The region in question is defined solely by its causal relationship to the conformal boundary of spacetime.  On the hypothesis that the causal structure of spacetime is primitive as argued elsewhere by Sorkin \cite{sorkin97}, and thus well defined even at the Planck scale, it seems reasonable to believe that a notion of region defined in terms of its causal relationships is likely to still make sense.
\begin{enumerate}
\setcounter{enumi}{\value{comment}}

\item All properties of $\mathcal{A}(t)$ are symmetric under time translation.  Thus each algebra $\mathcal{A}(t)$ is canonically isomorphic to the algebra at any one time, e.g.
$\mathcal{A}(0)$.

\setcounter{comment}{\value{enumi}}
\end{enumerate}
\noindent
Because time translation symmetry is used as an assumption, the proof applies only to a 1-parameter family of time slices on the horizon---a special case of the full GSL.
\begin{enumerate}
\setcounter{enumi}{\value{comment}}

\item The algebras $\mathcal{A}(t)$ are all contained as subalgebras of one big algebra $\mathcal{H}$, in such a way that each algebra also contains as a proper subalgebra all of the algebras in its future.

\setcounter{comment}{\value{enumi}}
\end{enumerate}
\noindent
$\mathcal{H}$ is the algebra of observables in the Heisenberg picture.  Each region $\mathcal{R}(t)$ contains the future regions, and therefore must contain all of the subregion's observables as a subalgebra.  Sorkin assumes that some information falls across the horizon and is lost, so that the algebras in $\mathcal{R}(t)$ do not include all observables from past times (cf. `Limitations' below for the results of dropping this assumption)

The structure defined above gives rise to the Schr\"{o}dinger time evolution, which is a positive linear trace-preserving map acting on the density matrices $\rho$ associated with $\mathcal{A}(0)$.  It is defined as follows: Although $\rho$ is in the statespace dual to $\mathcal{A}(0)$, by restriction $\rho$ may also be viewed as a state dual to the algebra at a later time $\mathcal{A}(t)$, $t > 0$.  One may then apply a backwards time-translation symmetry to the algebra $\mathcal{A}(t)$ in order to translate it into the algebra $\mathcal{A}(0)$, which transforms $\rho$ into a new state $\rho^\prime$.  This evolution is autonomous in the sense that it requires no information besides $\rho$ to calculate $\rho^\prime$.
\begin{enumerate}
\setcounter{enumi}{\value{comment}}

\item There exists a conserved energy operator $\hat{E}$ in $\mathcal{H}$ which is defined by the value of the fields at asymptotic infinity.  Because $\hat{E}$ is defined at infinity, it is always measurable outside the horizon and is therefore included in each algebra $\mathcal{A}(t)$.

\setcounter{comment}{\value{enumi}}
\end{enumerate}
\noindent
It follows from this that the Schr\"{o}dinger evolution also conserves energy.
\begin{enumerate}
\setcounter{enumi}{\value{comment}}

\item The space of states dual to $\mathcal{A}(0)$ has a finite number of states below any given energy $E_{max}$.
\end{enumerate}
\noindent
This assumption can only be true if the system has been placed in a box, e.g. AdS boundary conditions.  The restriction implies that every superselection sector of an algebra $\mathcal{A}(t)$ is described by a hyperfinite type I algebra (i.e. it is isomorphic to the algebra of all operators on some countable-dimension Hilbert space.)

Assumptions 1-5 plus the extra condition that there is only one superselection sector are enough to prove the GSL.  The microcanonical ensemble at any energy level $E$ is given by $\rho = 1/N$, where the natural number $N$ is the degeneracy of that energy level.  Sorkin begins by proving that this microcanonical ensemble evolves to itself as follows:  Consider the projection operator $\hat{P} = \delta (\hat{E},\,E)$ in $\mathcal{H}$ which projects onto the energy value $E$.  Since energy is conserved, $\hat{P}$ is also contained in $\mathcal{A}(t)$ for any value of t.  The microcanonical ensemble $\rho$ is defined in terms of $\hat{P}$ using the formula
\begin{equation}\label{trace}
\langle a \rangle_{\rho} = \mathrm{tr}(a\hat{P}/N)
\end{equation}
for any operator $a$ in $\mathcal{A}(t)$.  Now a single factor\footnote{The requirement of a single superselection sector is a hidden assumption of the proof not clearly stated in Ref. \cite{sorkin86}.  If there are multiple superselection sectors, it is easy to construct examples in which the maximum entropy state does not evolve to itself: e.g. three classical states A, B, and C where A and B evolve to A while C evolves to itself under time evolution.} of type I (or II) has a unique faithful normal semifinite trace\footnote{Some definitions: The trace of an operator algebra is defined as a positive linear function of algebra elements satisfying $\mathrm{tr}(AB) = \mathrm{tr}(BA)$ for all elements $A$ and $B$ in the algebra.  Semifinite means that every projection operator with infinite trace is the sum of two nonzero projection operators one of which has finite trace.  Normal means that the trace of an infinite sum of positive elements is equal to the sum of their traces.  A faithful trace is one that assigns a nonzero value to every projection operator but zero.} up to rescaling \cite{KR}.  Since the trace is unique, it does not matter whether Eq. (\ref{trace}) is defined using the algebra at time $t$ or the algebra at any previous previous time $t^\prime < t$.  As a result, the microcanonical ensemble is time-independent, i.e. it evolves to itself under time evolution.  Theorem 1 then shows that the outside entropy $S_{out}$ associated with any system of energy $E$ is nondecreasing.  Furthermore, by taking the sum of the microcanonical ensembles at all energies up to some $E_{max}$, one may invoke Theorem 1 to show that the entropy is conserved for any state with bounded maximum energy.  Since every normalizable state can be arbitrarily well-approximated by a state with sufficiently high maximum energy, continuity implies that all states exhibit entropy increase.

\paragraph{Limitations}

Unfortunately, these five assumptions, all of which are taken from Ref. \cite{sorkin86}, are mutually inconsistent.  For suppose that there were a set of algebras $\mathcal{A}(t)$ and $\mathcal{H}$ satisfying all of the above assumptions.  Let $\hat{Q}$ be the projection operator which projects onto states with energy $E > E_{max}$.  Restrict $\mathcal{A}(t)$ and $\mathcal{H}$ to the subalgebra of elements $a$ satisfying
\begin{equation}
\hat{Q}a = a\hat{Q} = 0,
\end{equation}
thereby obtaining the algebra of observables associated with the black hole system under the assumption that the energy is less than $E_{max}$.  These algebras $\mathcal{A_Q}(t)$ and $\mathcal{H_Q}$ are finite dimensional by virtue of assumption 5, and satisfy assumptions 2 and 4 by construction.  They also satisfy by construction assumption 3---except possibly for the criterion that each algebra be a \emph{proper} subalgebra of the future algebras, since it might be true that states with energy less than $E_{max}$ evolve by unitary evolution.  However, since assumption 3 requires that information loss occur for the complete algebras $\mathcal{A}(t)$, and since every normalizable state is arbitrarily close to one bounded by a sufficiently large energy bound, as long as $E_{max}$ is taken to be large enough the algebras $\mathcal{A_Q}(t)$ also satisfy assumption 3.  This implies that $\mathcal{A_Q}(1)$ is a proper subalgebra of any algebra $\mathcal{A_Q}(0)$.  But every proper subalgebra of a finite dimensional algebra has smaller dimension, so $\mathcal{A_Q}(1)$ has smaller dimension than $\mathcal{A_Q}(0)$.  This contradicts assumption 2 which states that the two algebras are isomorphic and therefore have equal dimension.

One possible way to bypass the contradiction is to deny assumption 5 by allowing there to be an infinite number of states below a given energy $E_{max}$.  There is then no contradiction since an infinite dimensional algebra can contain proper subalgebras isomorphic to itself.  To adapt Sorkin's proof it would be necessary to use one of the generalizations of Theorem 1 to the infinite dimensional case, which are given in Ref. \cite{lindblad75}.  One would need to show that there exists an equilibrium state and that despite the infinite dimensionality of the algebra, the nondecreasing quantity can still be reasonably identified with the finite Bekenstein-Hawking entropy of the black hole.

Another choice would be to keep the algebras $\mathcal{A}(t)$ finite-dimensional below any energy, but deny assumption 3 by permitting new degrees of freedom to be created near the black hole horizon to compensate for those degrees of freedom lost by falling into the black hole.  If this is the case, then the Heisenberg algebra $\mathcal{H}$ becomes infinite dimensional even though each algebra $\mathcal{A}(t)$ is finite dimensional.  The above method for obtaining the Schr\"{o}dinger time evolution would fail because the algebras $\mathcal{A}(t)$ would no longer be subalgebras of one another.  The positive linear trace-preserving map specifying the dynamics would depend on the details of how the new degrees of freedom entered the system.  Hence it is no longer possible to prove that the microcanonical ensemble evolves to itself, so additional assumptions are still needed.

Alternatively, one might drop the demand of assumption 3 by hypothesizing that the algebras $\mathcal{A}(t)$ are actually improper subalgebras of one another.  The observables outside the horizon would then evolve by a unitary evolution.  This would resolve the contradiction.  Also, one could immediately conclude from unitarity alone that the uniform probability state evolves to itself.  Since unitary evolution is a special case of a positive trace-preserving linear map the theorem would immediately show that $S_{out}$ is nondecreasing.  On the other hand, the entropy would also be nonincreasing unless some notion of coarse-graining were introduced.  The proof of the GSL would then become similar to proving the OSL (cf. section \ref{QG}).

\subsection{Semiclassical Quasi-Steady Version}\label{semi}

Sorkin has also proposed a similar proof applying in the semiclassical quasi-steady limit \cite{sorkin98}.  Rather than using the microcanonical ensemble, Sorkin now uses the ``Hartle-Hawking state''.  When restricted to the region outside both the black and white horizons of an eternal stationary black hole, this state is thermal with respect to the energy $E_{out}$ measured by a fiducial observer co-rotating just outside of the horizon.  There should be a generalized entropy associated with every spatial slice that terminates on the horizon.  Consider a family of such time slices $\Sigma(t)$ corresponding to the $t = \mathrm{const.}$ slices of some coordinate $t$ in which the background metric is time independent.  The state of this slice is then given by a density matrix $\rho$.  The generalized entropy is the sum of $A/4$ with $S_{out}$, the latter term being given by some renormalized version of the formula $-\mathrm{tr}(\rho \ln \rho)$.  Now if $t > 0$, all the information contained in the slice $\Sigma(t)$ is also contained in the slice $\Sigma(0)$, which means that $\rho(0)$ is sufficient to determine $\rho(t)$.  The evolution of $\rho$ from one time to another is therefore given by a positive linear trace-preserving map.  Actually, because the time evolution results from unitary time evolution followed by restriction, the map satisfies a stronger assumption known as complete positivity \cite{lindblad75}.\footnote{Complete positivity states that if the map acts on a system $A$ which is entangled with another independent system $B$, the resulting change in the combined system $AB$ also has the positivity property, i.e. positive states always evolve to other positive states.} 

In this setting the GSL states that the completely-positive time evolution map cannot decrease the generalized entropy.  Since the stationary state is a canonical ensemble, it does not assign to all states equal probabilities.  Sorkin uses a generalization of Theorem 1 to cover this case (a proof can be found in Ref. \cite{lindblad75}).

Theorem 2: \emph{Consider a quantum system described by the algebra of bounded operators on a countable-dimension Hilbert space (i.e. a type I hyperfinite von Neumann algebra), and a completely-positive trace-preserving linear map on the space of density matrices.  If the state which is thermal at temperature $T$ with respect to some ``energy'' operator $\hat{E}$ evolves to itself, then the free energy $\langle \hat{E} \rangle - TS$ of any initial state whatsoever cannot increase under this same evolution.}

Sorkin chooses $\hat{E}$ to be the fiducial energy outside the black hole horizon.  Applying Theorem 2 to the exterior of the semiclassical black hole, the change in $S_{out}$ over time is restricted by an inequality:
\begin{equation}\label{theorem}
\Delta(S_{out} - T^{-1}\langle E_{out})\rangle \ge 0.
\end{equation}
The semiclassical approximation allows Sorkin to equate the change in the black hole energy to the expectation value of the energy flowing into it.  Furthermore, the quasi-steady assumption that the flow of energy into the hole is uniform and slow permits one to ignore the time-profile of the response of the black hole to perturbations, and assume that the energy instantaneously increases the energy of the black hole, using the First Law of black hole thermodynamics (\ref{first}):
\begin{equation}\label{1st}
dS_{BH} = T^{-1}dE_{BH}.
\end{equation}
Combining (\ref{theorem}) with (\ref{1st}) gives
\begin{equation}
d(S_{BH} + S_{out}) \ge 0,
\end{equation}
which is the GSL.
\paragraph{Limitations}
Sorkin's approach seems to be very promising, but there are some gaps that still need to be filled before it can be regarded as a complete proof.

One problem is that the Hartle-Hawking state is not well-defined for black holes with superradiant modes.  This includes rotating black holes except when they are placed in a sufficiently small reflecting box \cite{DO08}.  The trouble is that there are field modes carrying a negative amount of fiducial energy, which makes the thermal state unnormalizable.  To get around this problem, the proof might need to be reformulated in a way that depends only on local events occurring near the horizon and not on global properties of the state.

A second issue needing resolution is the nature of the renormalization scheme used to define the entropy and energy.  As Sorkin says: 
\begin{quote}\small
It should be added that the matter entropy $S(\hat{\rho})$ we have been working with is actually infinite, due to the entanglement between values of the quantum fields just inside and just outside the horizon [...] Thus making our proof rigorous would require showing that changes in [Eq. \ref{theorem}] are nevertheless well-defined and conform to the temporal monotonicity we derived for that quantity. This probably could be done by introducing a high-frequency cutoff on the Hilbert space (using as high a frequency as needed in any given situation) and showing that he evolution of $\hat{\rho}$ remained unaffected because the high-frequency modes remained unexcited.  [From footnote (emphasis added):] \emph{In order to make the proof rigorous, one would also have, for example, to specify an observable algebra for the exterior fields and a representation of that algebra in which the operators $\hat{\rho}$ and $\hat{E}$ were well-defined (which in particular might raise the issue of boundary conditions near the horizon)}  (\cite{sorkin98}, p. 16)
\end{quote}

Thirdly, the above proof contains an unjustified assumption.  It is true that if one restricts the Hartle-Hawking state to a spatial slice $\Sigma$ bounded by the bifurcation surface one obtains a state thermal with respect to the Killing energy.  But if the slice $\Sigma$ passes through any other place on the horizon besides the bifurcation surface, it is not so obvious that the state is thermal.  Indeed, since a thermal state is normally defined using a notion of unitary time-translation symmetry, and since states on $\Sigma$ have no automorphisms generated by timelike Killing fields except when $\Sigma$ passes through the bifurcation surface, it is unclear what it would even mean to say that the state was thermal.

Since every faithful state is thermal with respect to some automorphism of the algebra of observables \cite{KR}, one might try to apply Theorem 2 to the free energy associated with this special automorphism of the restricted Hartle-Hawking state (known as the ``modular flow'').  Generically, the algebras of observables in bounded regions are expected to be type III von Neumann algebras, meaning that they do not have a trace at all.  This makes it difficult to define the free energy using the formula $\langle \hat{E} \rangle - TS$.  But rather remarkably, there exists a generalization of this concept of free energy to the context of an arbitrary von Neumann algebera, known as the ``relative entropy'' $S(\rho_{1} | \rho_{2})$ between two states $\rho_{1}$ and $\rho_{2}$.  This relationship is an asymmetrical one: if $\rho_{1}$ is regarded as a thermal state, $S(\rho_{1} | \rho_{2})$ can be thought of as the free energy of $\rho_{2}$ \cite{araki75}.\footnote{In some conventions the roles of $\rho_{1}$ and $\rho_{2}$ are reversed.}  Furthermore, Uhlmann \cite{lindblad75} has proven that the relative entropy is always nonincreasing when one restricts both $\rho_{1}$ and $\rho_{2}$ to a subalgebra, a result which may help prove the GSL.  However, the concept of the relative entropy is not always identical to the free energy defined by using the stress-energy tensor.  So it is still necessary to justify the use of the First Law (\ref{1st}) when the energy used is the modular flow.  Perhaps this could be done by taking some sort of near-horizon limit.

If these problems can be addressed, this proof promises to be of greater applicability than proofs using S-matrix techniques because the method allows one to discuss changes in the entropy of the black hole over a finite period of time.  This opens up the possibility that by replacing Eq. (\ref{1st}) with a more local formula like Eq. (\ref{response}) relating the stress-energy to the growth in area of a rapidly changing black hole, the quasi-steady assumption may be lifted.  The framework of slices also has the advantage over the S-matrix proofs that it is applicable to a black hole system contained in a reflecting box.

There are some more worrisome features, however, about attempting to extend this proof beyond the semiclassical domain.  The trouble is that the canonical ensemble is unnormalizable when the entropy of the black hole is taken into account, because the entropy increases faster than linearly with the energy.  This means that the Hartle-Hawking state is actually unstable.  If the black hole happens to grow a little, its temperature decreases and it continues to absorb more and more energy from its surroundings without limit.  If the black hole shrinks a little, its temperature increases and it evaporates more and more.  However, the timescale of the exponential growth is of order $R^3$ in Planck units.  Also, if the black hole is in equilibrium with a spherical ball of thermal radiation with radius greater than about $R^2$, the ball of radiation is itself unstable under collapse to a black hole over timescales of order $R^2$.  But since the semiclassical limit requires $R \gg 1$, neither of these instabilities can invalidate Sorkin's proof as applied to timescales of order $R$, the light-crossing distance.

\subsection{Combined with the S-matrix Approach}\label{comb}

Mukohyama \cite{muko97} has proven the GSL in a way that combines Sorkin's method using a time independent state with the S-matrix approach of Frolov \& Page (section \ref{Smat}).  This proof is a mathematically detailed form of Sorkin's argument applicable to any finite excitations of a free, real, massless scalar field on a quasi-steady collapsing black hole background.

The S-matrix for the scalar field on a stationary black hole background is a positive trace-preserving linear map going from the space of IN states to the space of OUT states.  Mukohyama begins by proving that if the IN state is in the canonical ensemble at the black hole temperature $T$ and angular velocity $\Omega$ (the Hartle-Hawking state), then the OUT state is also thermal at temperature $T$.  This implies that the free energy is nonincreasing when the same trace-preserving linear map is applied to any finitely excited IN state falling into the black hole (proven in Theorem 7 of Ref. \cite{muko97}).  The theorem only applies when the IN modes have a finite number of excitations above vacuum, despite the fact that the thermal state used to prove the theorem has infinitely many excitations.  Finally the First Law \ref{first} is used, as in section \ref{Smat}, to show the GSL.

\paragraph{Limitations}

The Hartle-Hawking state is ill-defined for superradiant black hole, yet it is used in an essential way in the framework of the proof.  As far as I can see, Mukohyama does not address this difficulty.

It would be nice if the proof could be generalized to more interesting forms of matter besides free massless scalar fields.  It would also be helpful to remove the requirement that the fields be finitely excited, because then the proof might be directly applicable to the thermal atmosphere of the black hole, which has infinitely many excitations (semiclassically) located closer and closer to the horizon.  In its current form the proof avoids directly analyzing the thermal atmosphere by using the S-matrix technique.

Because Mukohyama's proof uses an S-matrix, it only applies to asymptotic states, so the GSL can only be proven over finite time intervals by assuming that the matter falling into the black hole is also quasi-steady.\footnote{In this respect Mukohyama's proof is the same situation as every other quasi-steady proof reviewed here.  Cf. section \ref{SC})}  This limit is in tension with the requirement that the infalling matter be a finite excitation of the vacuum, but presumably this apparent contradiction can be reconciled by taking the quasi-steady limit of the infalling matter after invoking Mukohyama's Theorem 7.

\section{Proofs via the Generalized Covariant Entropy Bound}\label{BousB}

Now I will present a very different family of proofs, which explore the relationship between the Bousso bound and the GSL in the hydrodynamic regime, outside of the quasi-stationary limit.

Suppose one has a spacelike 2-surface $B$ from which a lightsurface $L$ emanates in one of the four possible lightlike and orthogonal directions.  Let the null rays on the lightsurface $L$ continue until terminating either on a cusp, a singularity, or a second spacelike boundary $B^{\prime}$.  If the null surface $L$ is initially nonexpanding at the surface $B$, and if the null energy condition holds on the horizon, then the area increase theorem shows that the $A^\prime$, the area of $B^\prime$, is always less than or equal to the area $A$ of $B$.  In this situation Flanagan, Marolf, and Wald (FMW) proposed a generalization of Bousso's covariant entropy bound (GCEB).  The GCEB states that the total entropy $S$ crossing the lightsurface $L$ is limited by the relation
\begin{equation}\label{GCEB}
S \le \frac{A - A^{\prime}}{4}.
\end{equation}
This bound---together with the null energy condition---immediately implies the GSL.  Simply take $B$ to be a slice of the horizon at one time, and $B^\prime$ to be a slice at an \emph{earlier} time.  (Since the light rays in $L$ are going backwards in time from $B$, the condition that the light rays are nonexpanding corresponds to the fact that the black hole's area is increasing with time).  So if one can prove equation (\ref{GCEB}) one also has a proof of the GSL.  The following two proofs do just this.\footnote{An additional argument for the Bousso bound not reviewed here is found in Ref. \cite{pesci07}}

In QFT entropy is not fully localizable, so the interpretation of $S$ in equation (\ref{GCEB}) is tricky.  The proofs below sidestep this nonlocality by explicitly using the hydrodynamic approximation, thus assuming that the entropy falling across $L$ is given by the integral of a fully localizable entropy current vector (cf. section \ref{hydro}).

\subsection{An Assumption Inspired by the Bekenstein Bound}\label{FMW}

The first proof of the GCEB was given by Flanagan, Marolf and Wald (FMW) \cite{FMW00}.  FMW assume that associated with every lightsurface $L$ there is an entropy current $s^a$ (thus $s^a$ might depend on the choice of $L$ as well as the spacetime coordinates).

FMW need to assume the following bound on $s^a$ in order to prove the GSL: Consider a generator of $L$, whose affine parameter is $\lambda$ at $B$ and whose tangent vector is defined as $k^{a} = (d/d\lambda)^{a}$.  This generator will either have infinite affine parameter length or else terminate at a finite affine parameter $\lambda^{\prime}$ when it hits the surface $B^{\prime}$, another generator in $L$, or perhaps a spacetime boundary such as a singularity.  If the generator goes on forever and is initially nonexpanding, then the null energy condition implies that $T_{ab}k^{a}k^{b} = 0$ along that generator, since any positive energy added to the right side of the Raychaudhuri equation (\ref{Ray}) would cause the generator to be trapped making it terminate at a finite value of the affine parameter.  In this case FMW assume that the entropy flux across the generator also vanishes.  If on the other hand the generator terminates, FMW restrict the entropy current $s^{a}_L$ flowing across the causal surface L to satisfy
\begin{equation}\label{assume1}
|s^{a}_{L}k_{a}| \le \pi(\lambda^{\prime} - \lambda)T_{ab}k^{a}k^{b}.
\end{equation}
According to FMW, ``the inequality [(\ref{assume1})] is a direct analogue of the original Bekenstein bound [(\ref{Bek})], with $|s^{a}_{L}k_{a}|$ playing the role of $S$, $T_{ab}k^{a}k^{b}$ playing the role of E, and [$\lambda^{\prime} - \lambda$] playing the role of $R$'' (\cite{FMW00} p. 4).  There are however a few differences between FMW's version and the original Bekenstein bound (\ref{Bek}).  In the original bound, $E$ refers to the time component of the total energy-momentum vector, and $R$ refers to an (orthogonal) spatial distance.  But FMW's bound relates the null energy to a null ``distance'' (this is invariant because both sides of Eq. (\ref{assume1}) transform the same way under a rescaling of the affine parameter).  More importantly, FMW's bound relates the local entropy density to the energy density instead of merely restricting the total amounts of both quantities.  This makes FMW's bound significantly more powerful than the original Bekenstein bound.  Furthermore, if the FMW bound is integrated in flat spacetime to relate the total null energy $E$ with the total entropy $S$, the numerical coefficient $\pi$ is a factor of two smaller than the coefficient $2\pi$ in the original Bekenstein bound (\ref{Bek}).  This also makes FMW's bound stronger than Bekenstein's bound.

I will now sketch FMW's proof.  In order to prove the GCEB (\ref{GCEB}), it is sufficient to show that it applies to each individual generator separately.  This can be shown trivially for generators of infinite affine length from FMW's assumption above that no entropy falls across infinite generators.  In the case of finite generators, the GCEB states that
\begin{equation}\label{I}
I \equiv \int_0^1 d\lambda\,s\mathcal{A}(\lambda)\,\le\,\frac{1}{4}[1 - \mathcal{A}(1)],
\end{equation}
where $s = -s_a k^a$ and the area-scaling factor is
\begin{equation}\label{defA}
\mathcal{A}(\lambda) = exp
\left[\int^{\lambda}_0 d\lambda^{\prime}\,\theta(\lambda^{\prime}) \right].
\end{equation}
Here FMW have used our freedom to rescale the affine parameter to make the integral go from 0 to 1 (if the affine parameter goes to infinity, then no entropy can cross it and the GCEB is automatically satisfied there).  The Raychaudhuri equation applied to the null generator says that
\begin{equation}\label{Raych}
-\frac{d\theta}{d\lambda} = \frac{1}{2}\theta^2 
+ \sigma_{ab}\sigma^{ab}
+ 8\pi T_{ab}k^ak^b,
\end{equation}
where $\sigma_{ab}$ is the shear tensor and the twist term is not included because null surfaces orthogonal to any boundary $B$ have vanishing twist.  FMW now define
$G(\lambda) = \sqrt{\mathcal{A}}$, and obtain from Eq's (\ref{defA}) and (\ref{Raych}) that
\begin{equation}\label{toG}
8\pi T_{ab}k^ak^b \le -2\frac{G^{\prime\prime}}{G}.
\end{equation}
Invoking the Bekenstein-like bound (\ref{assume1}), they obtain that
\begin{equation}\label{Q}
|s| \le (1 - \lambda)\pi T_{ab}k^a k^b.
\end{equation}
Substituting Eq. (\ref{Q}) into Eq. (\ref{I}) gives
\begin{equation}
I \le \int_0^1 d\lambda\,(1 - \lambda)\pi T_{ab}k^a k^b G^2.
\end{equation}
Eq. (\ref{toG}) can be used to re-express the integral as
\begin{equation}
I \le - \int_0^1 d\lambda\,(1 -\lambda)G^{\prime\prime}G/4.
\end{equation}
Since $0 \le G(\lambda) \le 1$ by the null energy condition, FMW drop it from the integrand and integrate the rest by parts:
\begin{equation}
I \le [G(0) - G(1) + G^{\prime}(0)]/4.
\end{equation}
Since $G(0) = 1$ by definition, $G(1) = \sqrt{\mathcal{A}(1)} \ge \mathcal{A}$, and $G^{\prime}(0) \le 0$ by the null energy condition, it follows that
\begin{equation}
I \le [1 - \mathcal{A}(1)]/4,
\end{equation}
which is the infinitesimal form of the Bousso bound as given in Eq. (\ref{I})  From this the GCEB and the GSL follow.

\paragraph{Limitations}

FMW's proof is valid outside the quasi-stationary limit, but they pay a price for it.  Not only must they assume the hydrodynamic approximation, the null energy condition, and few enough species for their Bekenstein-like bound to hold, but there are additional difficulties arising due to the difficulty of satisfying FMW's Bekenstein-like assumption (\ref{assume1}) over very short distances.

One must be careful in applying the Bekenstein Bound (\ref{Bek}) in the hydrodynamic approximation, because the bound is always violated by any nonzero entropy current in sufficiently small regions.  Both the entropy and the energy scale as the volume for constant density, causing the right side of (\ref{Bek}) to vanish faster than the left side.  This violation is an artifact of going beyond the validity of the hydrodynamic regime, since at sufficiently small distance scales the entropy is not as localizable as a classical current (cf. section \ref{hydro}).  Even quantum mechanics by itself is not sufficient to resolve this paradox, since in QM the entropy of independent subsystems is subadditive, which only makes the conflict with (\ref{Bek}) in small regions worse.\footnote{I believe that a proper understanding of the Bekenstein bound and entropy localization requires QFT considerations.  Because the entanglement entropy of field excitations makes the entropy diverge in any region with sharply defined boundaries, it is necessary to renormalize by somehow subtracting off the infinite entanglement entropy contribution from the vacuum to obtain a finite value for the entropy.  But since the entanglement entropy term being subtracted is itself subadditive, the resulting renormalized entropy can be superadditive whenever the entanglement entropy in the reference state used for subtraction exceeds the entanglement of the state being considered.  Consequently, it is possible to have the amount of entropy stored in a system be greater than the sum of the entropy of the parts.  This might permit something like a renormalized-Bekenstein bound to hold at all distance scales.}

Because the Bekenstein bound does not play well with the hydrodynamic regime, a fixed entropy current will always lead to violations of Eq. (\ref{assume1}) when one tries to apply the hydrodynamic limit outside of its scope.  For example, Eq. (\ref{assume1}) will not apply to a spherically symmetric star collapsing into a black hole, if one takes $B$ to be a slice of the horizon very close to its moment of formation, since whatever the finite ratio is between the entropy and energy at the center of the star when the horizon forms, $\lambda^{\prime} - \lambda$ can be taken to be small enough to violate Eq. (\ref{assume1}), despite the fact that the Bousso bound is just fine there.

This is why FMW's proof permits the entropy current to depend on the choice of $L$ as well as on the spacetime point---otherwise there are no nontrivial spacetimes in which Eq. (\ref{assume1}) is satisfied everywhere.  This is justified by FMW on the grounds that ``the entropy flux, $|s^{a}_{L}k_{a}|$, depends upon $L$ in the sense (described above) that modes that only partially pass through $L$ prior to [$\lambda^\prime$] do not contribute to the entropy flux'' (\cite{FMW00} p. 4).  However, permitting the entropy current to depend arbitrarily on $L$ is somewhat ad hoc.  It would be more elegant if the entropy currents associated with different choices of $L$ could be derived from a single common description of the matter flowing through the spacetime.

An alternative way to justify the entropy current's dependence on $L$ is given in Ref. \cite{BFM03}.  Violations of Eq. (\ref{assume1}) take place at small distance scales in which the hydrodynamic approximation is invalid.  So one may arbitrarily reconfigure the entropy current as long as the averages of the entropy current remain approximately constant at distance scales in which the hydrodynamic regime should be valid, in order to avoid violating $\ref{assume1}$ for a particular choice of $L$.  After all, the entropy current at distances smaller than the hydrodynamic regime is nonphysical anyway, so why not adjust its value to be most convenient?

\subsection{An Entropy Gradient Assumption}\label{BFM}

FMW also gave another proof of the (non-generalized) Bousso bound from different assumptions: namely a bound on the density and gradient of the entropy current, viewed as a vector on the spacetime independent of the choice of $L$.  This second proof does not yield the GSL because it only proves the ordinary Bousso bound.  In order to show that this set of assumptions could not lead to a proof of the GCEB, Guedens constructed an explicit counterexample to the \emph{generalized} Bousso bound given any fixed nonzero entropy current on spacetime \cite{guedens}.  In this example the GCEB (\ref{GCEB}) can be violated if $B$ is taken to be a 2-surface whose expansion parameter vanishes and $B^\prime$ is sufficiently close to $B$.  This violation occurs because the change in area is a quadratic function of the affine parameter interval $\Delta \lambda$, while the flux of entropy is a linear function of $\Delta \lambda$.  That means that the initial area change is not enough to satisfy Eq. (\ref{GCEB}) unless the entropy flux vanishes initially.  Consequently no proof of the GCEB is possible for all possible causal surfaces and fixed $s^a$.

Because of the counterexample, Bousso, Flanagan, and Marolf (BFM) \cite{BFM03} have constructed a modified proof which only tries to prove the Bousso bound for those causal surfaces which have no entropy falling across them initially.  As a bonus, this permits them to weaken the assumptions of Ref. \cite{BFM03}: they only need to restrict the gradient of the entropy, not the density.  Also, the numerical coefficient of the entropy gradient restriction is improved.

BFM assume the existence of an entropy current $s^{a}_{L}$ satisfying the following bound:
\begin{equation}\label{grad}
|s^{\prime}| \le 2\pi T_{ab}k^a k^b,
\end{equation}
where $s^{\prime} = -k^a k^b \nabla_{a}s_{b}$ and $k^a$ is the null vector generating the causal surface.  Note that Eq. (\ref{grad}) implies the null energy condition.  BFM also assume the isolation condition:
\begin{equation}\label{isolation}
s_{|B} = 0.
\end{equation}
They now attempt to prove that
\begin{equation}\label{noI}
\int_0^1 d\lambda\,s\mathcal{A}(\lambda)\,\le\,\frac{1}{4}[1 - \mathcal{A}(1)],
\end{equation}
which is the the GCEB as applied to an individual generator as given by Eq. (\ref{I}).  BFM obtain Eq. (\ref{toG}) again:
\begin{equation}
8\pi T_{ab}k^ak^b \le -2\frac{G^{\prime\prime}}{G},
\end{equation}
using the same argument given above.  From the gradient assumption (\ref{grad}),
\begin{equation}
s^{\prime}(\lambda) \le -\frac{G^{\prime\prime}(\lambda)}{2G(\lambda)}.
\end{equation}
Using the isolation assumption, BFM integrate the above assumption over $\lambda$ in order to bound the entropy density:
\begin{equation}
s(\lambda) \le -\int^{\lambda}_0 d\bar{\lambda}\,
\frac{G^{\prime\prime}(\bar{\lambda})}{2G(\bar{\lambda})}.
\end{equation}
Integrate this by parts:
\begin{equation}
s(\lambda) \le \frac{1}{2}\left[
\frac{G^{\prime}(0)}{G(0)} - \frac{G^{\prime}(\lambda)}{G(\lambda)}
- \int^{\lambda}_0 d\bar{\lambda}\,
\frac{G^{\prime}(\bar{\lambda})^2}{G(\bar{\lambda})^2}
\right].
\end{equation}
The first term is nonpositive when the causal surface is initially nonexpanding, and the third term is explicitly nonpositive.  Consequently these terms can be removed from the inequality:
\begin{equation}
s(\lambda) \le -{{G^{\prime}(\lambda)}\over{2G(\lambda)}}.
\end{equation}
BFM insert this inequality into the left-hand side of Eq. (\ref{noI}) and use $\mathcal{A} = G^2$:
\begin{equation}
\int_0^1 d\lambda\,s\mathcal{A}(\lambda) \,\le\,
-{1 \over 2}\int_0^1 d\lambda\, G(\lambda)G^{\prime}(\lambda) \,=\,
{1 \over 4}[G(0)^2 - G(1)^2].
\end{equation}
Since $G(0) = 1$ and $G(1)^2 = \mathcal{A}$, BFM obtain Eq. (\ref{noI}), proving the GCEB.

\paragraph{Limitations}

BFM make two different suggestions regarding how to interpret the isolation condition (\ref{isolation}) \cite{BFM03}.  One possible interpretation is that the condition restricts which lightsheets $L$ the proof is applicable to.  But then one would not be able to prove that generalized entropy increases from a time slice $\Sigma$ to a later time slice $\Sigma^{\prime}$, except when no entropy is falling into the horizon at time $\Sigma^{\prime}$.  Under that interpretation the GSL would not always follow from this proof.

Another suggestion is that rather than being a restriction on which causal surface may be considered, one should change the entropy current depending on the lightsheet $L$.  This would be similar to BFM's interpretation of the entropy bound described in the last paragraph of section \ref{FMW}.  One simply adjusts slightly the position of the entropy over small distance scales outside the validity of the hydrodynamic regime, to automatically satisfy the isolation condition.  This pushes all of the meaningful physical content into the gradient assumption (\ref{grad}) and the null energy condition, making it possible to prove the GSL for a much wider class of black hole horizon.

Why is there so much ambiguity in the interpretation of these proofs?  The hydrodynamic regime is at fault.  The trouble is the entropy current contains too much unphysical information even in those situations where a hydrodynamic approximation is appropriate.  Fixing this might require going beyond the hydrodynamic limit, or perhaps more carefully describing how to get a hydrodynamic entropy current from an actual state of matter.

\subsection{Weakening the Assumptions}\label{weak}

The assumptions (\ref{assume1}) and (\ref{grad}) can be weakened in two ways without compromising the ability to prove the GSL.  First of all one may replace $T_{ab} k^a k^b$ with $T_{ab} + \sigma_{ab}\sigma^{ab}/8\pi$ in the assumption and still use it to prove the GCEB, because the shear term is also present in the Raychaudhuri equation (\ref{Raych}) alongside the stress-energy term.  This additional term can thus be consistently interpreted as an ($L$ dependent) gravitational energy term which is added to the matter energy.  FMW consider adding in this extra term, saying ``we can then interpret $s^{a}_{L}$ as being the combined matter and gravitational entropy flux, rather than just the matter entropy flux'' (\cite{FMW00} footnote p. 4).  Since entropy stored in matter and entropy stored in gravitational radiation can be interconverted by means of ordinary thermal processes occurring away from any black holes, it seems inevitable that the outside entropy term used when defining the GSL must include gravitational entropy.  So the best version of this proof probably includes the shear term.

Secondly, the absolute value signs in assumptions (\ref{assume1}) or (\ref{grad}) are also unnecessary for proving the GSL.  Thus one may replace them with the assertion that each generator of $L$ with finite affine length satisfies either
\begin{equation}\label{weak1}
s \le (\lambda^{\prime} - \lambda)(\pi T_{ab}k^{a}k^{b} + \sigma_{ab}\sigma^{ab}/8),
\end{equation}
or else
\begin{equation}\label{weak2}
s^{\prime} \le (2\pi T_{ab} k^a k^b + \sigma_{ab}\sigma^{ab}/4).
\end{equation}
Similarly, if the affine parameter is infinite, then instead of requiring $s = 0$ in the first proof one only needs to require $s \le 0$.  The weakening of this assumption only makes a difference in situations when $s$ is negative which requires that $s^a$ be spacelike or null.  However, these assumptions are not sufficient to prove the GCEB because the GCEB counts positively all the entropy that crosses the causal surface $L$ regardless of the direction of the entropy flow.

As an example of a situation in which one might want to assign a negative $s$, consider a black hole which is radiating Hawking quanta outward but which is kept critically illuminated by incoming pure matter.  Since entropy is being radiated from the horizon, a hydrodynamic description of the system requires the entropy flowing into the horizon to be negative.  Admittedly, this situation is probably outside the hydrodynamic regime's validity.  But as long as the entropy current on the horizon is a good approximation to the change in $S_{out}$ over time, the approximation is sufficient for purposes of proving the GSL.  It does not matter if the entropy current is unphysical in other respects.

Strominger and Thompson (ST) \cite{ST04} have pointed out that in BFM's proof, the isolation condition (\ref{isolation}), the condition that the lightsheet $L$ be initially nonexpanding, and the null energy condition can all be replaced with a single, weaker condition:
\begin{equation}
s_{|B} \le -\theta/4.
\end{equation}
The proof then essentially states that if the GSL is satisfied at $B$, it is satisfied on the entire causal surface.  This is more elegant than the seemingly arbitrary conditions of BFM's proof.  It also helps to explain why the GSL should apply to global event horizons, which are defined using a nonlocal ``teleological'' boundary condition.  According to this modified proof, one can prove that a generator of a causal surface satisfies the GSL only so long as it also satisfies the GSL at any later time.  This can be phrased in a more local way by saying that every generator which begins to violate the GSL cannot ever change back into a generator which satisfies the GSL.

In the same paper ST propose that the GSL beyond the hydrodynamic regime is related to a quantum-corrected version of Bousso's covariant entropy bound, in which the entanglement entropy is added to the area.  Unfortunately they are not able to make this provocative conjecture precise except in the two-dimensional RST model.  ST give a proof of the quantum Bousso bound in this setting, but it only applies when the matter is in a coherent state.

In the following section I will discuss a proof of the GSL for coherent states in this RST model, by Fiola, Preskill, Strominger, and Trivedi \cite{fiola94}.  However, unlike the ST's proposed quantum Bousso Bound, the proof in the next section applies to the apparent horizon, rather than to the event horizon (cf. \ref{horizon}).

\section{2D Black Holes}\label{2D}

Since it is hard to analyze important questions of quantum gravity in 3+1 dimensions, it might well be more tractable to first consider the analogous issues in 1+1 dimensions.  The 1+1 Einstein-Hilbert action is topological field theory, and therefore has no local degrees of freedom.  However, one may reintroduce local degrees of freedom by adding a scalar field to produce ``dilaton gravity'' \cite{fiola94}.  There are many different possible actions one can write down for this scalar field.  Many of the resulting theories are equivalent to restricting to just the s-wave sector in a higher dimensional theory.

There exists a 1+1 dimensional model, found by Russo, Susskind, and Thorlacius (RST), which is exactly solvable in the large $N$ limit and yet also includes finite backreaction effects due to Hawking radiation.  One does this by taking the limit that Planck's constant $\hbar$ goes to zero while holding $N \hbar$ fixed so that the backreaction due to Hawking radiation remains finite.  The hope is to prove the GSL in regimes beyond the quasi-stationary limit by means of an exact calculation.  Because this proof is based more on calculation than on conceptual analysis, it is specific to the RST model.  Therefore, I will first present the RST model, and then go on to describe the proof of the GSL for coherent states in this model.

\subsection{The RST model}

RST \cite{RST92} began with the action of the classical CGHS model \cite{CGHS91}:
\begin{equation}
\mathcal{S}_{classical} = \frac{1}{2\pi}\int d^2x \sqrt{-g}
\left[ e^{-2\phi}(R + 4(\nabla \phi)^2 +4\lambda^2))
- \frac{1}{2}(\nabla_{\mu} f_i \nabla^{\mu} f_i) \right].
\end{equation}
Here $g$ is the determinant of the metric, $R$ is the curvature scalar, $\phi$ is the dilaton field, $f_i$ are the $N$ scalar fields, and the repeated index $i$ is summed over.  In black hole like solutions, the value of the dilaton varies over the spacetime in such a way that the theory is weakly coupled far from the black hole and strongly coupled inside near the ``singularity''.  Null coordinates $x^+$ and $x^-$ may be defined having the property that
\begin{equation}
g_{++} = g_{--} = 0.
\end{equation}
The event horizon is the boundary which separates the outgoing light rays that escape to the weakly coupled region from the outgoing light rays that fall into the strongly coupled region.   On the other hand, the apparent horizon is located where $\partial_{+}\phi$ vanishes.  These two definitions of the horizon agree for a stationary black hole.  Let $\phi_H$ represent the value of $\phi$ on the horizon.  One may then calculate in the usual ways the mass:
\begin{equation}\label{CGHS}
M_{BH} = \frac{\lambda}{\pi}e^{-2\phi_H},
\end{equation}
the temperature (which is independent of the mass):
\begin{equation}
T_{BH} = \frac{\lambda}{2\pi},
\end{equation}
and the entropy:
\begin{equation}\label{2dBH}
S_{BH} = 2e^{-2\phi_H}.
\end{equation}
(These properties all agree with those for a near-extremal magnetically charged black hole in 4 dimensional dilaton gravity \cite{GM88}, a theory which reduces to the CGHS model when restricted to classical s-waves.)

There are semiclassical correct corrections to the theory even in the large $N$ limit.  Fluctuations in the metric and dilaton are negligible, and the corrections to the stress energy of the scalars $f_i$ can be calculated using the conformal anomaly.  The one loop correction is equivalent to a classical theory with a nonlocal term added to the action of Eq. (\ref{CGHS}):
\begin{equation}
\mathcal{S}_{loop} = -\frac{N}{96\pi}\int d^2x \sqrt{-g(x)} \int d^2y \sqrt{-g(y)}R(x)G(x,y)R(y),
\end{equation}
where $G(x,y)$ is the Green's function of $\nabla^2$.  Adding an additional counterterm of the form
\begin{equation}
\mathcal{S}_{counter} = -\frac{N}{48\pi} \int d^2x \sqrt{-g} \phi R,
\end{equation}
makes the resulting RST model is exactly solvable.  Defining $\rho$ implicitly by means of the nonzero component of the metric in null coordinates as follows:
\begin{equation}\label{null}
g_{+-} = -e^{2\rho}/2,
\end{equation}
and redefining the fields so that
\begin{equation}\label{Omega}
\Omega = \frac{12}{N}e^{-2\phi} + \frac{\phi}{2} + \frac{1}{4}\ln \frac{N}{48},
\end{equation}
and
\begin{equation}
\chi = \frac{12}{N}e^{-2\phi} + \rho - \frac{\phi}{2} - \frac{1}{4}\ln \frac{N}{3},
\end{equation}
the action 
$\mathcal{S}_{eff} = \mathcal{S}_{classical} + \mathcal{S}_{loop} + \mathcal{S}_{counter}$
takes the form:
\begin{equation}
\mathcal{S}_{eff} = \frac{1}{\pi} \int d^2x \left[
\frac{N}{12}(-\partial_{+}\chi\, \partial_{-}\chi + \partial_{-}\Omega\, \partial_{+}\Omega
+ \lambda^2 e^{2\chi - 2\Omega} + \frac{1}{2} \partial_{+}f_i\, \partial_{-}f_i \right]
\end{equation}
The scalar fields $f_i$ are now decoupled from $\Omega$ and $\chi$.  Further simplification comes by choosing the null coordinates $x^{+}$ and $x^{-}$ so that the relation
\begin{equation}\label{Kruskal}
\chi = \Omega,  
\end{equation}
which is equivalent to
\begin{equation}
\rho = \phi + \frac{1}{2}\ln\frac{N}{12},\end{equation}
holds on-shell.  This is one way of fixing the parameter $\rho$ in Eq. (\ref{null}), which makes the exact solubility manifest.  Another choice is the sigma coordinates (also defined only on-shell) which are related to the null coordinates as follows:
\begin{equation}
\lambda x^{+} = e^{\lambda \sigma^{+}},\quad \lambda x^{-} = -e^{-\lambda \sigma^{-}}.
\end{equation}
These $\sigma$ asymptotically correspond to the inertial coordinates at $\mathcal{I}^{-}$, which means that the vacuum built on them is the state that contains no quanta as measured by asymptotic observers to the past.

$\Omega$ is not a monotonic function of $\phi$; rather, it has a minimum at a critical value:
\begin{equation}
\phi_{cr} = -\frac{1}{2}\ln \frac{N}{48},\quad \Omega_{cr} = \frac{1}{4}.
\end{equation}
Values of $\Omega$ less than $\Omega_{cr}$ do not correspond to any value of $\phi$ and are therefore unphysical.  So wherever the fields reach the critical value actually corresponds to a boundary of the spacetime.  When this boundary is timelike, the RST model requires reflecting boundary conditions in order to be complete.  This corresponds to the ``origin'' of spacetime in the 3+1 dimensional analogue.  When this boundary is spacelike, it corresponds to the singularity of the 3+1 dimensional black hole---and in fact, it is a curvature singularity in 1+1 dimensions as well.  Strong coupling occurs where $\Omega \sim \Omega_{cr}$, near the origin or the singularity, while weak coupling occurs when $\Omega \gg \Omega_{cr}$, far from the black hole.

\subsection{The Entropy Formula}

According to the abstract of Fiola, Preskill, Strominger, and Trivedi (henceforth FPST) \cite{fiola94}, ``a generalized second law of thermodynamics is formulated, and shown to be valid under suitable conditions.''  One of these conditions is that the matter falling upon the black hole must be in a coherent state.  FPST state that if the infalling matter is not coherent, then sometimes the GSL is violated.  This claim, if true, would be even more remarkable than the proof itself.  However, some of the assumptions behind this claim are questionable, such as FPST's formula for the total entropy, and the choice of the apparent horizon over the event horizon for defining the GSL.  I will begin by discussing these assumptions, and then will go on to cover their proof.

The generalized entropy should be a number associated with any spacelike slice terminating on a point on the horizon.  FPST proposed formula is:
\begin{equation}\label{three}
S_{tot} = S_{BH} + S_{BO} + S_{FG},
\end{equation}
where $S_{BH}$ is the entropy of the black hole itself (which classically is given by Eq. (\ref{2dBH}), $S_{FG}$ represents the entanglement entropy of the quantum fields outside the black hole, and $S_{BO}$ is associated with the entropy of the matter falling into the black hole.  FPST evaluate Eq. (\ref{three}) on the \emph{apparent} horizon.

\subsubsection{The Fine-Grained Entropy}

\begin{figure}[ht]
\centering
\includegraphics{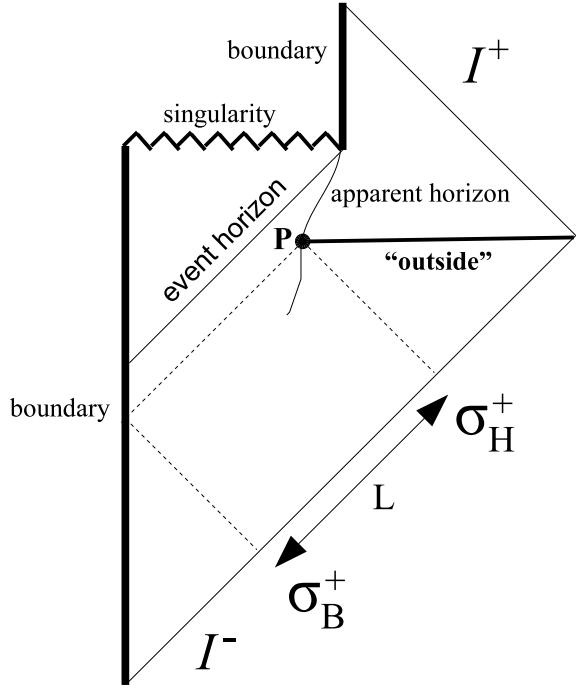}
\caption{A Penrose diagram of the two dimensional black hole.  The point $P$ on the apparent horizon can be traced backwards to $\sigma^{+}_B$ or $\sigma^{+}_H$.  The ``outside'' is the region whose fine-grained entropy is being calculated.}\label{2Dfig}
\end{figure}

$S_{FG}$, the ``fine-grained'' entropy, is calculated by considering the entanglement entropy outside of the horizon, when the fields are in a vacuum state with respect to the $\sigma$ coordinates (i.e. with respect to inertial observers at $\mathcal{I}^{-}$).  It is the Gibbs entropy $-\mathrm{tr}(\rho\ln \rho$) when one restricts this state to the system outside of the horizon.  Before giving its formula FPST need to define some auxiliary variables.  Given a point $P$ on the apparent horizon, there are two possible lightlike directions going backwards in time (see Figure \ref{2Dfig}).  One way goes straight to $\mathcal{I}^{-}$ at $\sigma^{+} = \sigma^{+}_H$, while the other reflects off the ``origin'' and then hits $\mathcal{I}^{-}$ at $\sigma^{+} = \sigma^{+}_B$.
FPST define $L = \sigma^{+}_H - \sigma^{+}_B$ as the difference between these coordinates.  They also need an ultraviolet cutoff at a proper distance $\delta$ from the horizon because the entanglement entropy is logarithmically divergent near the horizon.  FPST can now calculate the result as
\begin{equation}\label{FG}
\frac{N}{6} \left[ \phi_{H} - \phi_{cr} + \frac{\lambda L}{2} + \ln \frac {L}{\delta} \right],
\end{equation}
up to an error of order unity which can be absorbed into $\delta$.  For technical reasons, FPST's calculation is only valid under the simplifying assumption that there is no infalling energy prior to $\sigma^{+}_B$ (matter falling in before then would make it impossible to simultaneously satisfy the Kruskal gauge given by (\ref{Kruskal}), and the equality between the $\sigma^{+}$ and $\sigma^{-}$ coordinates on the reflecting boundary prior to the formation of the black hole).  As the point $P$ approaches the point of final evaportation, $\sigma^{+}_B$ limits to the moment at which the event horizon forms.  Consequently, to validate (\ref{FG}) everywhere on the horizon, FPST must assume that no matter falls into the black hole prior to the formation of the event horizon.

Any coherent state of a free field has field expectation values given by a classical solution, and quantum fluctuations around the mean field values of exactly the same magnitude as in the vacuum state.  Since the shift in expectation values makes no difference to the entanglement entropy, the exact same formula (\ref{FG}) can be used whenever the incoming matter takes the form of a coherent state built on the $\sigma$ vacuum (so long as there is no infalling matter falling in prior to the time $\sigma^{+}_B$, as stated above).

\subsubsection{The Black Hole Entropy}

$S_{BH}$, the entropy of the black hole, is classically just given by Eq. (\ref{2dBH}), but there are quantum corrections.  FPST calculate this by considering a black hole in a box in equilibrium with its radiation.  By inserting a little bit of energy into the black hole from outside and using the First Law, they can calculate $\Delta S_{BH} + \Delta S_{FG}$ of the entire system.  This, however, causes the black hole to grow and consume some of the outside radiation, so $\Delta S_{FG}$ must be subtracted off in order to find the total change in $\Delta S_{BH}$.  This then yields $\Delta S_{BH}$ up to a constant, which FPST fix by requiring the black hole to have zero entropy when it reaches zero size (that is, when $\phi_{H} = \phi_{cr} = (1/2)\ln (N/48)$  The result is
\begin{equation}\label{BH}
S_{BH} = 2e^{-2\phi_H} - \frac{N}{12}\phi_H - 
\frac{N}{24}\left[ 1 + \ln \left( \frac{N}{24} \right) \right].\footnote{For some reason this term does not agree with the black hole entropy calculated by Myers \cite{myers94}, using Wald's Noether charge method.}
\end{equation}

Note that the formula above does not depend on the value of the horizon cutoff $\delta$, whereas the formula for $S_{FG}$ given by (\ref{FG}) does.  This means that the total fine-grained entropy $S_{BH} + S_{FG}$ of a given state depends on the cutoff $\delta$.  This result is paradoxical because $\delta$ should ultimately be taken to zero (at least semiclassically), which would make the entropy of the black hole diverge.  However, the dependence of the generalized entropy on $\delta$ is only an additive constant in the two-dimensional case, meaning that it cancels out when calculating changes in the entropy.  As FPST say, ``the sensitivity to the cutoff does not prevent us from making definite statements about how the entropy outside the black hole \emph{changes} during its evolution, or about the change in the intrinsic entropy of the black hole itself'' (\cite{fiola94} p. 4006).  There is no problem since FPST are only interested in comparing two times when the horizon is present.  However, the $\delta$ dependence does not cancel out when comparing a time with a horizon to a time without a horizon, or in higher than two dimensions.  So checking that the GSL holds at the instant of formation or collapse, or performing a similar analysis in more than 2 dimensions, would require some sort of renormalization procedure (cf. section \ref{SC})

\subsubsection{The Boltzmann Entropy}\label{Boltz}

The final term $S_{BO}$, the Boltzmann entropy, is intended to take into account the entropy of the matter falling into the black hole.  Recall that FPST restrict their consideration to states in which the infalling matter is in a coherent state.  Coherent states are always pure.  In the Gibbs point of view, a pure state must be assigned zero entropy, yet a robust proof of the GSL requires that matter with nontrivial entropy be allowed to impinge upon the hole.  FPST tell us that ``even though the incoming matter is in a pure state, it surely carries thermodynamic entropy.  We can assign a nonzero entropy to this state by performing a coarse-graining procedure'' (\cite{fiola94} p. 4006).  In other words, they wish to use the Boltzmann entropy for defining the entropy of the infalling matter while retaining the Gibbs picture for the outgoing Hawking radiation.  The infalling matter has a left-moving energy profile:
\begin{equation}
\mathcal{E}(\sigma^{+}) \equiv \frac{12\pi}{N} T_{++}(\sigma^{+}),
\end{equation}
using the same unconventional normalization of $\mathcal{E}$ as FPST.  FPST treat $\mathcal{E}$ as a measurable macroscopic observer, and assign to it an entropy based on the logarithm of the number of states of left-movers with the same energy profile.  They calculate this to be
\begin{equation}\label{BO}
S_{BO} = \frac{N}{6} \int_{\Sigma_{out}} d\sigma^{+} \sqrt{\mathcal{E}(\sigma^{+})}.
\end{equation}
As the coherent excitation falls into the black hole, $S_{BO}$ can only decrease over time.  This means that the addition of the $S_{BO}$ term only makes it harder to satisfy the GSL.

I believe that this approach to calculating the entropy of infalling matter is problematic.  In the Boltzmann picture a coarse-graining procedure is only justified if the information being ignored is somehow irrelevant to the evolution of the system.  This might be the case if the microstate is in some sense a typical member of the macrostate in question, or if all members of the macrostate evolve in an indistinguishable way at the microscopic level.  Neither condition is satisfied here because most pure states are not coherent, and coherence is necessary for the calculation of the value of $S_{FG}$ as given by Eq. (\ref{FG}).  In other words, the coherent state is not a typical member of its macrostate class.

On the other hand in the Gibbs perspective, this step involves the unwarranted substitution of a mixed state for the pure incoming state.  Either one retains the pure state, in which case the entropy of the incoming matter is zero, or else one considers a bona fide incoherent mixed state, in which case there is no guarantee that (\ref{FG}) is valid.  As FPST themselves admit:
\begin{quote}\small
While the expression [(\ref{three})] may appear (and indeed, is) somewhat strange, we believe it to be a precise two-dimensional analogue of the notion of `total entropy' used implicitly in discussions of four-dimensional black hole thermodynamics.  This prescription might be interpreted as follows.  We may consider, instead of a pure initial state, the mixed initial state $\rho$ that maximizes $-tr \rho\ln \rho$ subject to the constraint that the energy density is given by the specified function $\mathcal{E}(\sigma^{+})$.  For this mixed initial state we have $S_{Boltz} = -tr \rho\ln \rho$.  What we are adding to $S_{BH}$ in [Eq. (\ref{three})] is the fine-grained entropy outside the horizon for this particular mixed initial state.  [Footnote (emphasis added):] \emph{Note that we have not really established that this interpretation is correct.  In particular, our expression for $S_{FG}$ has been derived only for coherent incoming states, and may not apply for arbitrary states.}  In any event we have not been able to find any other reasonable and precise alternative to [Eq. (\ref{three})] that obeys a generalized second law.  (\cite{fiola94} p. 4007)
\end{quote}
Additionally, even if $S_{BO}$ were the correct formula for the infalling entropy far from the horizon, one must take into account the ``observer dependence'' \cite{MMR04} of the entropy---the fact that the entropy attributable to an object depends not only on the object but also on how close it is to the horizon of the observer measuring its entropy.  Thus a system with a given entropy at spatial infinity will have a different entropy when it is lowered down to just outside a black hole event horizon.  The reason is that the system is now sitting on top of the black hole's thermal atmosphere, whose entropy it raises less than it would have raised the vacuum.  This means that $S_{BO}$ and $S_{FG}$ cannot simply be added together.

A more defensible prescription for the generalized entropy is $S_{BH} + S_{out}$, where $S_{out} = -tr \rho\ln \rho$ of the region outside of the horizon at the time being considered.  This formula has no need to distinguish which component of the entropy is due to the entanglement and which component is due to the matter; it is simply the total fine-grained entropy of the region.  However, it requires the specification of a renormalization procedure to be valid (cf. section \ref{SC}).

\subsection{Which Horizon?}\label{horizon}

Is it correct to use the global event horizon or the apparent horizon for purposes of the GSL?  The choice makes a significant difference outside of the quasi-steady limit.  The usual opinion is that one ought to use the event horizon.  However, FPST take a contrary view:
\begin{quote}\small
We find it more appropriate to define $S_{BO}$, $S_{FG}$ and $S_{BH}$ using the apparent horizon, for several reasons.  First of all, the position of the apparent horizon can be determined locally in time, without any required information about the global properties of the spacetime.  Our observer on a time slice can readily identify the apparent horizon as the location where $\partial_{+} \Omega$ vanishes.  Second, because the position of the apparent horizon is determined by this local condition, it is easy to compute the trajectory of the apparent horizon using the RST equations.  (\cite{fiola94} p. 4006)
\end{quote}
These reasons are not very convincing.  The fact that the location of the event horizon is sensitive to nonlocal considerations does not by itself amount to an argument that it cannot be a physically relevant concept.  Concepts relying on global structure (such as the notion of thermal equilibrium in QFT) are often quite important to physics.  Furthermore, there is no reason why a concept of physical interest should also be easy to calculate in a given model.  FPST continue:
\begin{quote}\small
Third, if we use the global horizon to define the entropy, the resulting thermodynamic expressions do not seem to have a nice thermodynamic interpretation.  In particular, the would-be second law is easily violated by sending in a very sharp pulse with a large entropy and energy density but small total entropy and energy.  The essential point is that the value of the dilaton at the global horizon responds less sensitively to the incoming pulse than does the dilaton at the apparent horizon.  (\cite{fiola94} p. 4007)
\end{quote}
Note that because the RST model is the s-wave sector of a 4 dimensional theory, this argument threatens to invalidate the use of the event horizon in general and not just in the two dimensional case.  This startling claim is not explicated further by FPST, so I will attempt to elucidate their argument further.  (I will describe the argument using the more familiar four dimensional black hole, whose entropy is the horizon area, since the essential features are the same in any dimension).  Suppose the infalling matter consists of a thin spherical shell containing energy $E$, entropy $S$, and proper radial length $r$, as measured far from the black hole.  If the shell is hurled at the speed of light into a black hole of radius $R$ at the speed of light, the event horizon will anticipate the shell by growing to nearly its final size before the shell even begins to cross the horizon.  The horizon finishes its growth when the shell has completely crossed the horizon.  Therefore, in the limit that $r \to 0$, the event horizon has already grown to its final area when the shell falls in.  But when the shell falls in it reduces the outside entropy by an amount equal to $S$, without any instantaneous change in $S_{BH}$.  Consequently the generalized entropy of the event horizon decreases when the shell crosses the horizon.  This violation would not apply to the apparent horizon because the apparent horizon does not anticipate the infall of matter but only grows while the shell is actually falling in.

But can $r$ can really be taken to zero while $E$ and $S$ are held fixed?  It is easy to show that the Bekenstein bound would forbid this limit, since (assuming the bound refers to the narrowest dimension of the shell), it would require that
\begin{equation}
S \le 2\pi rE.
\end{equation}
Now if $E$ and $r$ are both small, the total change in horizon area, over the interval that the shell falls through, is proportional to $rE$, which is greater than $S$ by virtue of the bound.  However, in the RST model the Bekenstein bound is violated parametrically due to the large numbers of species.  So if the generalized entropy is given by Eq. (\ref{three}), the GSL can be violated for the event horizon by sending in a thin shell containing many species and thus large $S_{BO}$.  This violation can be seen as an additional reason to reject Eq. (\ref{three}) beyond those given in section \ref{Boltz}.

Suppose that instead of using Eq. (\ref{BO}), one asks how much fine-grained entropy the shell adds to the thermal atmosphere of the black hole.  When the shell is a distance $r$ from the black hole horizon, every part of it is immersed in a thermal bath of temperature greater than or equal to $1/{2\pi r}$.  Assuming the shell's energy is a small perturbation to the thermal atmosphere, the Clausius relation says that
\begin{equation}\label{Beklike}
\Delta S \le 2\pi r\Delta E.
\end{equation}
So even though the Bekenstein bound does not hold for isolated objects containing large numbers of species, when the objects are close to the horizon of the black hole, the quantity $\Delta S$ does satisfy a bound with the same form as the Bekenstein bound.  So if the Bekenstein bound prevents violations of the GSL, Eq. (\ref{Beklike}) prevents GSL violations even in the case of large $N$.  So the event horizon may well obey the GSL in FPST's thin-shell thought experiment.  However, since the above argument is dimensional, it can only establish that no parametric violation of the GSL occurs.  Conceivably, a violation could still be present if the factors of order unity work out badly.  Since the situation goes beyond both the quasi-steady and hydrodynamic regimes, it is outside of the scope of any of the sound arguments included in this review.

There is yet another reason to prefer the event horizon to the apparent horizon: the GSL can be violated otherwise.  This is demonstrated in Appendix B of FPST's paper, which shows that for noncoherent states, the generalized entropy given by (\ref{three}), as applied to the apparent horizon, can temporarily go down.  FPST say how:
\begin{quote}\small
[...] quantum states can be constructed that pack a large positive density of (fine-grained) entropy without carrying a large energy density.  We can prepare matter in such a state, and allow the matter to fall into a black hole.  Then the fine-grained entropy decreases sharply, but without any compensating sharp increase in the black hole entropy.  Hence the total entropy decreases.

Alternatively, we can make the total entropy decrease (momentarily) by simply sending in negative energy into the black hole.  It can be arranged that the black hole shrinks and loses entropy without a compensating increase in the fine-grained entropy.  (\cite{fiola94} p. 4012)
\end{quote}
The remainder of their Appendix is devoted to constructing such states by choosing an alternative vacuum defined using a function of the $\sigma^{+}$ coordinate.  FPST construct the analogue of the formula for the fine-grained entropy (\ref{FG}) which is valid for this new vacuum state, and show that the total entropy as given by (\ref{three}) can be made to temporarily decrease.  It is well-known that negative energy densities can be made to exist for short periods or small regions in QFT, so long as they are balanced by even greater positive energies elsewhere, whose size is governed by certain ``quantum inequalities'' \cite{FR95}.  The negative energy density between two conducting plates due to the Casimir effect are an example.  If such negative energy densities fall across the horizon of a black hole, the apparent horizon will instantly decrease in size and thus lose entropy.  The only way to prevent GSL violation would be if the entanglement entropy in the negative energy region always increases enough to compensate.  FPST explicitly calculate $S_{FG}$ to show that this does not occur for certain negative energy density pulses in the RST model.  It may be shown in the case of the Casimir energy by a simple scaling argument: As the distance $x$ between the Casimir plates decreases, the energy density scales like $x^{-d}$ where $d$ is the spacetime dimension, while any finite change in the entanglement entropy across a slice going between the plates scales like $x^{2-d}$.

I have argued above that the formula $S_{BH} + S_{FG} + S_{BO}$ is incorrect, but it is not the problem here.  FPST have calculated $S_{FG}$ in the vacuum state with respect to any choice of null coordinate, and dropping the Boltzmann entropy term does not resolve the GSL violation.  The problem is the choice of the apparent horizon, which responds instantly to any negative energy perturbation.  Whereas the event horizon can expand even when negative energy falls into it, so long as the negative energy will be followed by positive energy of sufficient magnitude and closeness in time.  (This property of the event horizon has already been shown by Ford and Roman \cite{FR01} to be necessary to save the GSL from the negative energy fluxes associated with non-minimally coupled scalar fields.)  Energy inequalities may therefore be important in determining whether the event horizon can violate the GSL beyond the quasi-steady limit.

\subsection{A Proof for Coherent States}

In summary, FPST have assumed so far that:
\begin{enumerate}
\item the system is described by the RST model,
\item the generalized entropy is given by $S_{tot} = S_{FG} + S_{BH} + S_{BO}$ on the apparent horizon, and
\item no energy falls into the black hole prior to the formation of the event horizon.
\end{enumerate}
They have also calculated each of the three terms in the generalized entropy.

The first step is to add up the expression $S_{FG} + S_{BH} + S_{BO}$ in order to obtain the total entropy.  They begin by adding the first two terms (\ref{FG}) and (\ref{BH}) together, and then using (\ref{Omega}) to re-express the result in terms of $\Omega$ instead of $\phi$.  The result is
\begin{equation}\label{two}
S_{BH} + S_{FG} = \frac{N}{6} \left[ 
\Omega_H - \frac{1}{4} + \frac{\lambda L}{2} + \ln \frac{L}{\delta}
\right].
\end{equation}
Next they solve for $\Omega_H$ based on the energy profile $\mathcal{E}$ of the infalling matter, using the definition of the apparent horizon $\partial_{+} \Omega = 0$ to obtain
\begin{equation}
\Omega_H = \frac{1}{4} + \frac{M}{\lambda} - \frac{\lambda L}{4},
\end{equation}
where $M$ is defined by
\begin{equation}
M(\sigma^{+}_H) = \int^{\sigma^{+}_H}_{-\infty} d\sigma^{+}\,\mathcal{E}(\sigma^{+}).
\end{equation}
Adding everything together including the Boltzmann entropy (\ref{BO}), the final result is
\begin{equation}
S_{total} = \frac{N}{6} \left[ 
\frac{1}{\lambda} M(\sigma^{+}_H) + \frac{\lambda L}{4} + \ln \frac{L}{\delta}
+ \int^{\infty}_{\sigma^{+}_H} d\sigma^{+}\,\sqrt{\mathcal{E}(\sigma^{+})}
\right].
\end{equation}
FPST now calculate that
\begin{equation}
\frac{\partial \sigma^{-}_H}{\partial \sigma^{+}_H} = 
e^{-\lambda L} \left( 1 - \frac{\mathcal{E}(\sigma^{+}_H)}{\mathcal{E}_{cr}} \right),
\end{equation}
where $\mathcal{E}_{cr}$ is the critical infalling energy needed to balance out the Hawking radiation to keep the size of the black hole constant.  Since
\begin{equation}
L = \sigma^{+}_H - \sigma^{+}_B = \sigma^{+}_H - \sigma^{-}_H + const.,
\end{equation}
the derivative of L is
\begin{equation}
\frac{\partial L}{\partial \sigma^{+}_H} = 
1 + e^{-\lambda L} \left( \frac{\mathcal{E}}{\mathcal{E}_{cr}} \right).
\end{equation}
This makes it possible to calculate the derivative of $S_{tot}$ in terms of 
$\tilde{\mathcal{E}} = \mathcal{E} / \mathcal{E}_{cr}$
as
\begin{equation}
\frac{\partial S_{tot}}{\partial \sigma^{+}_H} =
\frac{N \lambda}{24} \left[
(\sqrt{\tilde{\mathcal{E}}(\sigma^{+}_H)} - 1)^2 + 
e^{-\lambda L} (\tilde{\mathcal{E}}(\sigma^{+}_H) - 1)
\left( 1 + \frac{4}{\lambda L} \right) + \frac{4}{\lambda L} \right].
\end{equation}
Although it is not exactly manifest, this formula is always positive when 
$\tilde{\mathcal{E}} \ge 0$ and $L > 0$.  Therefore the GSL is established given the above assumptions.  Unfortunately, because the result comes from a calculation rather than a conceptual proof, the reason for the increase in entropy is mysterious and may be model dependent.

\section{Prospects}

A summary of the proofs can be found in the Table of Proofs.  The table indicates the authors, information about the the regime (cf. section \ref{regimes}), as well as what extra assumptions or problems there are.  Although there are many proofs, the only ones that appear to be completely sound are Hawking's area theorem (\cite{hawking71} section \ref{class}), the three proofs in the hydrodynamic regime (\cite{wald94} section \ref{atm}, \cite{FMW00}\cite{BFM03} section \ref{BousB}), and Frolov and Page's proof from the S-matrix (\cite{FP93} section \ref{Smat}).  However the conceptual foundations of the hydrodynamic approximation are not completely clear, and it may be that hydrodynamic proofs are only valid in the classical regime.

A natural next step would be to attempt a proof of the GSL in the semiclassical but non-quasi-steady regime.  A strategy for constructing such a proof would be to take a semiclassical quasi-steady proof and find a way to remove the quasi-steady assumption.  Such a proof would have to take into consideration the the nontrivial response of the event horizon's area to the infalling energy profile, which is described by Eq. (\ref{response}).  This could be used to generalize to a new regime not covered by the semiclassical quasi-steady proofs of Frolov and Page \cite{FP93} (section \ref{Smat}), Sorkin \cite{sorkin98} (section \ref{semi}), or Mukohyama \cite{muko97} (section \ref{comb}).

Because the GSL involves assertions about the increase of generalized entropy on arbitrary time slices of the black hole spacetime, the S-matrix approach of Frolov and Page's proof seems to be highly dependent on the quasi-steady limit to ensure that what happens in the asymptotic past and future is relevant for proving the GSL at finite times.  Sorkin's semiclassical proof is a more likely starting point, because the theorem used in the proof allows one to make deductions about the entropy difference between any two time slices.  Although for technical reasons this proof is invalid, if the problem can be fixed, it may well also lead to important results outside the quasi-steady limit.

An alternative strategy would begin with one of the non-quasi-stationary hydrodynamic proofs and try to promote it to a proof valid in the semiclassical limit.  Here Strominger and Thompson's proposal \cite{ST04} for generalizing the Bousso bound to a fully quantum setting by adding the entanglement entropy to the area seems to be promising (cf. section \ref{weak}).  Since the weaker version of the Bousso bound was important for formulating the GCEB which implied the GSL in the hydrodynamic regime, it stands to reason that this quantum-corrected Bousso bound might be used to show the GSL in the semiclassical setting.  However, for it to help with proving the GSL in higher dimensions, this quantum-corrected Bousso bound must first be formulated and proven in dimensions higher than two.  Even in two dimensions the proof of the bound is so far limited to coherent states in the RST model.  It might be best to start by proving the bound in more general two-dimensional situations, perhaps by adapting one of the more general proof methods.  (Although two-dimensional proofs like that of FPST \cite{fiola94} (section \ref{2D}) are attractive because some two-dimensional models are exactly solvable, their downside is that any proof which takes advantage of an exact solution must necessarily be limited to particular models.)

In order to proceed with either of these two strategies, a more rigorous approach to the renormalization of $S_{out}$ is probably needed.  Because the entropy diverges near the horizon, one naive renormalization procedure is to put a membrane $M$ just outside the black hole event horizon, and find the entropy outside of the membrane $M$.  Then one might hope to renormalize this entropy while taking the limit that $M$ approaches the horizon.  Finally one would have to show that all of the different ways of taking this limit give the same result.  However, this procedure fails because $M$ is a perfectly sharp boundary which is itself associated with an infinite entanglement entropy.

Instead, one might use the mutual information, defined as the difference between the sum of the entropy of two systems and the entropy of the combination of both the systems (in other words, the mutual information measures the extent to which the entropy of a system is less than the sum of the entropies of its parts).  The mutual information between the region inside the event horizon and the region outside of $M$ should be finite so long as there is a finite proper distance between every point on $M$ and the horizon \cite{CM04}.  Other possible ways to regularize the entropy divergence are given in Ref. \cite{BKLS86}.

Another approach would be to try to frame the proof of the GSL using algebraic QFT.  If the generalized entropy can be defined directly in terms of the infinite algebra associated with the region outside of the event horizon, then it may be possible to entirely sidestep any need to renormalize a finite entropy.

Another mystery of the GSL as presently formulated is why it applies to the event horizon, which is teleologically defined in terms of what is going to happen in the future.  However, the ultimate proof of the GSL must be framed entirely within a theory of quantum gravity.  If the GSL is ultimately true because of quantum gravitational physics occurring at the Planck scale, it seems a little strange that it should only apply to event horizons and not to all causal surfaces whatsoever.  But some causal surfaces disobey the GSL, as discussed in section \ref{choice}.  So it would be nice if some local principle could be found which applies to all causal surfaces and which implies the GSL for event horizons.  Such a principle might be provable using only the physics close to the horizon.  Perhaps then, by having a theory of generalized thermodynamics broad enough to apply to all causal surfaces everywhere, it will be easier to see what features a microscopic theory of quantum gravity needs in order to give rise to macroscopic thermal behavior.

\vspace{-4pt}

\small
\subsubsection*{Acknowledgements}

This work was supported in part by NSF grant PHY-0601800, the Perimeter Institute, and the Maryland Center for Fundamental Physics.  I would like to thank Rafael Sorkin and Rob Myers for discussing their work with me, and my advisor Ted Jacobson for extensive comments.

\begin{landscape}
\begin{center}
TABLE OF PROOFS\phantom{spacingitoutalittle}
\end{center}
\begin{tabular*}{1.4\textwidth}{@{\extracolsep{\fill}}|l|c|c|c|c|}
  \hline
\footnotesize{PROOF} & \footnotesize{REGIME} & \footnotesize{PERTURB.} & 
\footnotesize{EXTRA CONDITIONS AND/OR DIFFICULTIES} & \footnotesize{SECTION} \\

  \hline
Hawking  \cite{hawking71}   &classical& any     &null energy condition, cosmic censorship     &\ref{class} \\
  \hline
Zurek \& Thorne  \cite{ZT85}& semi.  & q-steady &entropy localization, renormalization &\ref{heur} \\
  \hline
Wald  \cite{wald94}         &hydro.  & q-steady &adiabaticity (fixable)  &\ref{atm} \\
  \hline
Frolov \& Page  \cite{FP93} & semi.  & q-steady &CPT insufficient for charged BH (fixable) &\ref{Smat} \\
  \hline
Sorkin 1  \cite{sorkin86}   & full QG& any      &inconsistent assumptions   &\ref{full} \\
  \hline
Sorkin 2  \cite{sorkin98}   & semi.  & q-steady &thermality, not superradiant, renormalization&\ref{semi} \\
  \hline
Mukohyama  \cite{muko97}    & semi.  & q-steady &not superradiant, free scalar field   &\ref{comb} \\
  \hline
Flanagan et al.  \cite{FMW00}&hydro.  & any      &null energy condition, Bekenstein-like bound &\ref{FMW} \\
  \hline
Bousso et al.  \cite{BFM03} &hydro.  & any      &entropy gradient bound, isolation condition  &\ref{BFM} \\
  \hline
Fiola et al.  \cite{fiola94} & semi.  & any      &RST model, large N, apparent horizon  &\ref{2D} \\
  \hline
\end{tabular*}
\end{landscape}

\end{document}